\documentclass[prb,twocolumn,showpacs]{revtex4}
\usepackage{graphicx}
\usepackage{amssymb}
\usepackage{dcolumn}
\usepackage{float}
\usepackage{bm}
\usepackage{booktabs}
\usepackage{amsmath} 
\usepackage[utf8]{inputenc}
\usepackage{lmodern}
\usepackage{booktabs}
\usepackage{color}
\usepackage{color}
\usepackage[breaklinks=true,colorlinks,citecolor=magenta,linkcolor=blue,urlcolor=magenta]{hyperref}

\newcolumntype{M}[1]{>{\centering\arraybackslash}m{#1}}

\newcommand{\bs}{\boldsymbol}
\DeclareMathAlphabet{\bi}{OML}{cmm}{b}{it}

\begin{document}


\title{Dynamical polarization and plasmons
in noncentrosymmetric metals}

\bigskip
\author{Sonu Verma, 
Arijit Kundu and Tarun Kanti Ghosh\\
\normalsize
Department of Physics, Indian Institute of Technology-Kanpur,
Kanpur-208 016, India}

\date{\today}
\begin{abstract}
We study the dynamical polarization function and plasmon modes for spin-orbit coupled noncentrosymmetric metals such as Li$_2$(Pd$_{1-x}$Pt$_x$)$_3$B. These systems have different Fermi surface topology for Fermi energies above and below the spin degenerate point which is also known as the band touching point (BTP). We calculate the exact dynamical polarization function numerically and also provide its analytical expression in the long wavelength limit. We obtain the plasmon dispersion within the framework of random phase approximation. In noncentrosymmetric metals, there is a finite energy gap in between intra and interband particle hole continuum for vanishing excitation wavevector. In the long wavelength limit, the width of interband particle hole continuum behaves differently for Fermi energies below and above the BTP as a clear signature of the Fermi surface topology change. 
We find a single undamped optical plasmon mode lying in between the intra and interband particle hole continuum for Fermi energies above and below the BTP within a range of parameters. The plasmon mode below the BTP has smaller velocity than that of above the BTP. It is interesting to find that as we tune the Fermi energy around the BTP, the plasmon mode becomes damped within a range of electron-electron interaction strength.  For Fermi energies above and below the BTP, we also obtain an approximate analytical result of plasma frequency and plasmon dispersion which match well with their numerical counterparts in the long wavelength limit. The plasmon dispersion is $\propto q^2$ with $q$ being the wave vector for plasmon excitation in the long wavelength limit. We find that varying the carrier density with fixed electron-electron interaction strength or vice versa does not change the number of undamped plasmon mode, although damped plasmon modes can be more in number for some values of these parameters. We demonstrate our results by calculating the loss function and optical conductivity which can be measured in experiments.
\end{abstract}
\maketitle

\section{Introduction}
For several decades, ubiquitous role of spin-orbit interactions \cite{Rash1,Rash2,Dress}
in various condensed matter systems \cite{spintro1,spintro2,spintro3} exhibiting 
exotic phenomena has been observed \cite{dissp1,dissp2,SHE1,SHE2,SHE3,SHE4,SHE5,SHE6,spin-torque,SGE1}. 
The charge carrier's spin is not a conserved quantity
in spin-orbit coupled systems, which facilitate to control the spin
by simply electric manipulation.
The study of response functions in presence of external perturbations 
in spin-orbit coupled systems with electron-electron interactions plays a 
vital role in understanding several fundamental many body properties of 
the systems. 
Single particle excitation spectra and the collective modes of the systems 
are determined by the dynamical response functions which incorporate the 
dynamical screening of Coulomb interaction\cite{giuliani,bruusflensberg}.
Whereas static response function govern the transport properties of the systems 
through the scattering by charge impurities in presence of screened Coulomb 
interaction\cite{giuliani,bruusflensberg}.
Also many body properties such as dielectric function and collective excitation 
spectrum of systems with spin-orbit interaction (SOI) have several importance 
in terms of understanding the many-body correlations and observation of SOI 
effects in these systems\cite{2dplasmon1,2dplasmon2,metaldichalconide,2dplasmon3,2dplasmon4}. 
Two dimensional electron-hole gas (2DEG/2DHG) with Rashba SOI (RSOI) and Dresselhaus SOI (DSOI) 
in a single quantum well host isotropic and anisotropic plasmon 
spectrum when considered one type of SOIs and both SOIs, respectively\cite{2dplasmon4,2DHGanisotropic}. 
Moreover, 2DEG with RSOI in double quantum well hosts both lower energy acoustic and optical 
plasmon modes with charge density oscillating out of phase and in phase in a neutralizing 
positive background\cite{2dplasmon2}. 


In recent years, there have been several theoretical and experimental studies on materials 
showing spin-orbit interaction much higher than that of semiconductor heterostructures. 
Examples of such materials are three-dimensional (3D) topological insulators \cite{BiSe1,BiSe2}, 
Bi/Ag(111) surface alloy\cite{Bi_Alloy}, 3D bipolar semiconductor BiTeX (X=Cl,Br,I) 
\cite{BiTeI1,BiTeI2,BiTeI3,BiTeI4,BiTeI5}. In BiTeX compounds both in bulk and surface, 
the giant RSOI arises due to the local electric field as a 
consequence of inversion asymmetry. 
According to ${\bf k} \cdot {\bf p}$ perturbation theory\cite{BiTeI2}, the RSOI in 
these materials have a planar form like $\alpha ({\bs \sigma}\times {\bf k})_{z}$ with $\alpha$ 
being the strength of RSOI, ${\bs \sigma}$ being a vector of spin Pauli matrices  
and ${\bf k}$ being electron's wave vector. 
In addition to BiTeX compounds, B20 \cite{B201}  compounds and noncentrosymmetric metals
such as Li$_2$(Pd$_{1-x}$Pt$_x$)$_3$B\cite{NC_Li1} also show strong RSOI due to lack of
inversion symmetry. The leading order SOI experienced by conduction electrons in these materials
is described by $\alpha {\bs \sigma}\cdot{\bf k}$, which is quite different from the bipolar
semiconductor compounds. 
These systems with strong RSOI possess a distinct property that the Fermi surface topology
changes as one tune the Fermi energy across the band touching point (BTP) of two
spin-split bands. It has been verified both theoretically and experimentally
\cite{susRashba,susprl} that the system changes its behavior from paramagnetic
to diamagnetic as Fermi energy sweeps across the BTP from below. There are also
several studies in BiTeX compounds 
\cite{Trans1,Trans2,Trans3,Trans4,Trans5,Trans6,Trans7,RKKY,Therm1,Therm2,Mag_pht,Opt3,op-cond-ti}
and noncentrosymmetric metals\cite{sup1,rev-sup,NC_Li2,NC_Li3,spin_scp,RKKY,thermNCMs} 
in the context of transport, magnetic, thermoelectric and optical response showing 
distinct behavior below and above the BTP due to change in the Fermi surface topology. 
All these electronic properties mainly based on the single particle excitations of the systems.
Moreover, collective modes in BiTeX compounds  have been studied thoroughly \cite{Opt3}.
The study of collective modes in noncentrosymmetric metals is still lacking. 
The focus of this paper is to look into several aspects of the 
charge collective modes of noncentrosymmetric metals by studying 
the full dynamical polarization function within the random phase 
approximation (RPA) in detail. 


In this work, we calculate the dynamical polarization function 
(also known as Lindhard function) 
numerically and also provide its analytical form for small $q$.  
The spin-orbit coupled systems possess intra and interband single 
particle hole continuum (PHC). Latter is also known as \textit{Rashba continuum}. In the long wavelength limit, the width of Rashba continuum responds to the change in the Fermi surface topology and shows different behavior for Fermi energies above and below the BTP. In noncentrosymmetric metals (NCMs), interband PHC starts at finite energy at $q=0$.  In presence 
of electron-electron interaction within the framework of jellium model, we calculate 
the plasmon dispersion within RPA. Due to isotropic nature of the band structure, we find a single 
optical undamped plasmon mode in between the intraband PHC and Rashba continuum 
within a range of material parameters of NCMs. In the long wavelength limit, we provide an approximate 
analytical formula for plasma frequency and plasmon dispersion. The plasmon dispersion is
$\propto q^2$ in the long wavelength limit similar to that of ordinary 3D electron gas\cite{giuliani}. The plasmon dispersion and plasma frequency extracted from 
both numerical and analytical results match well for small $q$. For Fermi energies below BTP, we find that the 
plasmon mode has smaller velocity than that of Fermi energies above BTP. This plasmon mode becomes damped for 
Fermi energies near the BTP due to the shift in the Rasbha continuum towards 
zero energy within a range of electron-electron interaction strength. We also find only 
one single undamped plasmon mode by varying the electron-electron interaction strength, 
although there are more number of plasmon modes lying within the Rashba continuum 
for a range of interaction strength for Fermi energies below and above the BTP. We calculate 
the loss function and optical conductivity within RPA to demonstrate the plasmon mode 
which can be observed in experiments.

Remainder of this paper is organized in the following manner. In Sec. II, the necessary 
ground state properties of NCMs are given. In Sec. III, we discuss the intra 
and interband PHC derived from the dynamical polarization function. The static
Lindhard function along with its singularities are also discussed.  
Section IV describes the plasmon dispersion in detail together with the energy loss function 
and optical conductivity which can be measured experimentally. 
We summarize our results in Sec. V.

\section{Ground state properties}
\begin{figure}[htbp]
	\begin{center}\leavevmode
		\includegraphics[width=0.45\textwidth]{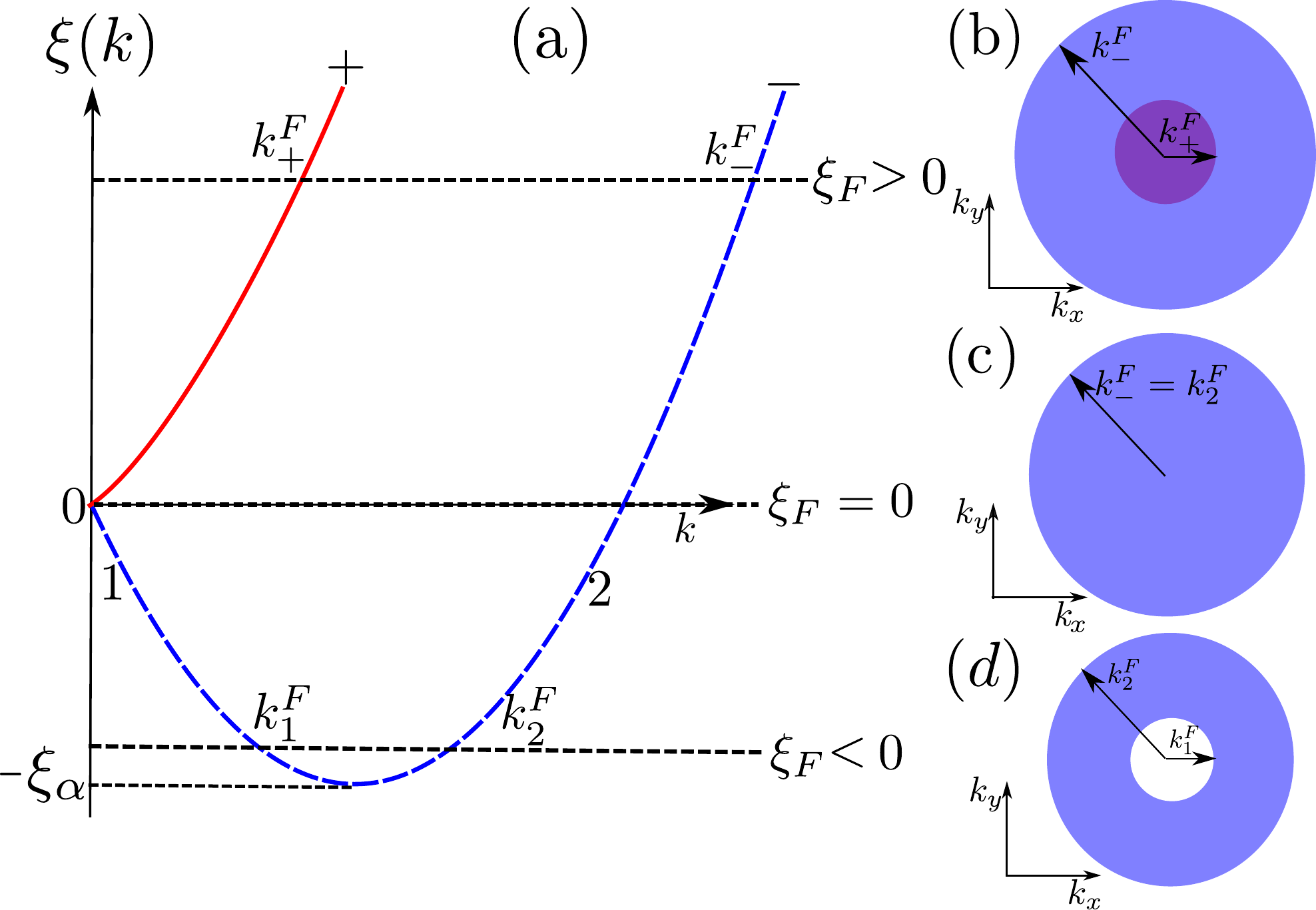}
		\caption{(a) Energy dispersion of noncentrosymmetric metals:
			The  $k=0$ point where two bands touch is known as band touching point (BTP).
			In panels (b) and (d) the cross-sections of the Fermi surfaces for $\xi_F > 0$ and $\xi_F < 0$ are shown, respectively.
			The Fermi surface topology is different in both the cases having convex-convex shape and concave-convex shape 
			for $\xi_F>0$ and  $\xi_F<0$, respectively. (c) There is only one Fermi surface at BTP ($\xi_F=0$) where the change in the Fermi surface topology occurs.}
		\label{sketch}
	\end{center}
\end{figure}

The low energy conduction electrons in a 3D noncentrosymmetric metal can be effectively 
described by the following non-interacting Hamiltonian near the $\Gamma$ point\cite{B201,NC_Li2,fullbandstruc} : $H=H_0 + H_D$ where
\begin{align}\label{Ham1}
H_0 =\frac{\hbar^2 {\bf k}^2}{2m^\ast} \sigma_0 + \alpha  \; {\bs \sigma} \cdot {\bf k},
\end{align}
and 
\begin{align*}
H_D=\beta [k_x\sigma_x(k_y^2-k_z^2)+k_y\sigma_y(k_z^2-k_x^2)+k_z\sigma_z(k_x^2-k_y^2)].
\end{align*}
Here $m^\ast$ is the effective mass of an electron, $\sigma_0$ is 
$2 \times 2 $ unit matrix, ${\bs \sigma} = \{\sigma_x,\sigma_y,\sigma_z\}$ 
is a vector of Pauli spin matrices, $ {\bf k} = \{k_x, k_y, k_z\}$ is the electron's 
wavevector, $\alpha $ characterizes the strength of the RSOI, and $\beta$ is the strength of cubic spin-orbit coupling term which breaks the $C_4$ symmetry. It has been argued that the presence of the cubic spin-orbit coupling term in the Hamiltonian does not change transport and magnetic properties qualitatively\cite{B201}. In this work we ignore the cubic spin-orbit coupling ($H_D$).
As helicity operator ${\bf k}\cdot{\bs \sigma}/k$ commutes  
with the Hamiltonian $H_0$, from now onwards we will work in the eigen basis of the helicity
operator having eigenvalues $\lambda = \pm 1$. 
Thus, the eigenstates of the 
above Hamiltonian will be 
$\psi_{{\bf k},\lambda}({\bf r})=\phi_{{\bf k},\lambda} e^{i {\bf k \cdot r}}/\sqrt{\mathcal{V}}$, 
where $\mathcal{V}$ is volume of the system, $\lambda = \pm 1$ represents two opposite helicities, 
and $\phi_{{\bf k},\lambda}$ is helicity eigenstate which takes the following forms:
\begin{align}
\phi_{{\bf k},+} & = 
\begin{bmatrix}
\cos(\theta/2) \\
e^{i\phi} \sin(\theta/2)
\end{bmatrix},
\hspace{0.3cm}
\phi_{{\bf k},-} & = \begin{bmatrix}
\sin(\theta/2) \\
- e^{i\phi} \cos(\theta/2)
\end{bmatrix}.
\end{align}
Here, $\theta$ and $\phi$ are the polar and azimuthal angles, respectively, 
which represent the orientation of ${\bf k}$. The energy dispersion consists 
of two spin-split bands corresponding to $\lambda = \pm  $ having the structure 
$\xi_{{\bf k},\lambda} = \hbar^2 k^2/(2m^\ast) + \lambda \alpha k$. Due to distinct 
spin-momentum locking, these systems have different Fermi surface topology for energy 
$\xi>0$ (convex-convex shape) and $\xi<0$ (concave-convex shape) as shown 
in Fig. \ref{sketch}. 
There are two Fermi wavevectors 
$k_{\lambda}^{F}= -\lambda k_{\alpha} + \sqrt{k_{\alpha}^{2} + 2m^*\xi_F/\hbar^2}$ with 
$k_{\alpha} = m^*\alpha/\hbar^2$, corresponding to $\lambda = \pm$ bands for $\xi_F>0$. 
The density of states  for $\lambda = \pm $ bands become
\begin{eqnarray}\label{dof1}
D_{\lambda}^{>}(\xi_F) & = & D_0  
\Bigg[\frac{\xi_F+2\xi_\alpha}{\sqrt{\xi_F+\xi_\alpha} }
-\lambda \sqrt{4\xi_\alpha} \Bigg],
\end{eqnarray}
where $D_0 = \frac{1}{4\pi^2} (\frac{2m^\ast}{\hbar^2} )^{\frac{3}{2}}$ and 
$\xi_{\alpha} = \hbar^2 k_{\alpha}^2/2m^*$.
The total density of states is given by 
$D^{>}(\xi_{F})= 2 D_0
\frac{(\xi_F+2\xi_\alpha)}{\sqrt{\xi_F+\xi_\alpha}} $.
For $\xi_F<0$, $\lambda = -$ band is characterized by the two branches with the Fermi wavevectors 
$k_{\eta}^{F}=  k_{\alpha} -(-1)^{\eta-1} \sqrt{k_{\alpha}^{2} + 2m^*\xi_F/\hbar^2}$ with $\eta=1,2$.
The density of states within two concentric spherical shells with radii $k_1$ and $k_2$ are given by
\begin{eqnarray}\label{dof2}
D_{\eta}^{<}(\xi_F) = D_0 
\Bigg[\frac{\xi_F+2\xi_\alpha}{\sqrt{\xi_F+\xi_\alpha} }
-(-1)^{\eta-1} \sqrt{4\xi_\alpha} \Bigg],
\end{eqnarray}
with total density of states $D^{<}(\xi_{F})= 2D_0 
\frac{(\xi_F+2\xi_\alpha)}{\sqrt{\xi_F+\xi_\alpha}} $. For $\xi<0$, the $\lambda =-$ band has a 
non-monotonic behaviour and has a van Hove singularity in the density of states at 
$\xi=-\xi_{\alpha}$ with $\xi_{\rm min} = -\xi_{\alpha}$, 
similar to the conventional 1D electron gas. \\
In the $T \rightarrow 0$ limit, the Fermi energy $\xi_F$ can be extracted from the 
following equation
\begin{eqnarray}\label{fermi}
(4 \xi_{\alpha} + \xi_F) \sqrt{\xi_{\alpha} + \xi_F} =  (\xi_{F}^{0})^{3/2},
\end{eqnarray} 
where $\xi_{F}^{0} = \frac{\hbar^2}{2m^*}(3\pi^2 n_e)^{2/3}$ is the Fermi energy 
for ordinary 3D electron gas with $n_e$ being the density of the conduction 
electrons in NCMs. 
It can be easily seen from above equation that $n_e =n_t$ with $ n_t = 4k_{\alpha}^3/3\pi^2$ 
is the critical  density of electrons where the Fermi surface topology changes which 
also defines the band touching point (BTP).\\

\section{Dynamical polarization function}
Within the linear response theory for translationally invariant systems, the 
dynamical polarization function or density-density correlation function of 
the two-level system in response to a time-dependent perturbation in Fourier space  
becomes (see Appendix \ref{app-pf})  
$\chi^{0}_{\rho\rho}({\bf q},\omega) = 
\sum_{\lambda\lambda^{\prime}}\chi^{0}_{\lambda\lambda^{\prime}}({\bf q},\omega)$, with
\begin{align}\label{chi0}
&\chi^{0}_{\lambda\lambda^{\prime}}({\bf q},\omega+i0^+) \nonumber\\
&=\sum_{\bf k}\frac{\mathcal{F}_{\lambda\lambda^{\prime}}({\bf k},{\bf k+q})}{\mathcal{V}} 
\frac{n^{F}_{{\bf k},\lambda} - n^{F}_{{\bf k+q},\lambda^{\prime}} }
{\hbar\Omega+\xi_{{\bf k},\lambda}-\xi_{{\bf k+q},\lambda^{\prime}}},
\end{align}
where $\hbar\Omega=\hbar(\omega+i0^+)$ and 
$n^{F}_{{\bf k},\lambda}=1/[e^{\beta(\xi_{{\bf k},\lambda}-\mu)}+1]$ 
with $\beta=(k_{\rm B} T)^{-1}$, $T$ being the temperature. 
Also $\mathcal{F}_{\lambda\lambda^{\prime}}({\bf k},{\bf k+q}) = 
|\phi^{\dagger}_{{\bf k},\lambda}\phi_{{\bf k +q},\lambda^{\prime}} |^2 = 
\frac{1}{2}[1+\lambda\lambda^{\prime}\frac{{\bf k}\cdot{\bf (k+q)}}{|{\bf k}| |{\bf k+q}|}]$ 
describes the overlap between the states labelled by 
$|{\bf k},\lambda\rangle$ and $|{\bf k+q},\lambda^{\prime}\rangle$. 
In the above notation of dynamical polarization function the subscript `$\rho\rho$'  indicates that it is a density-density correlation function.
\begin{figure*}[htbp]
\begin{center}\leavevmode
\includegraphics[width=0.9\textwidth]{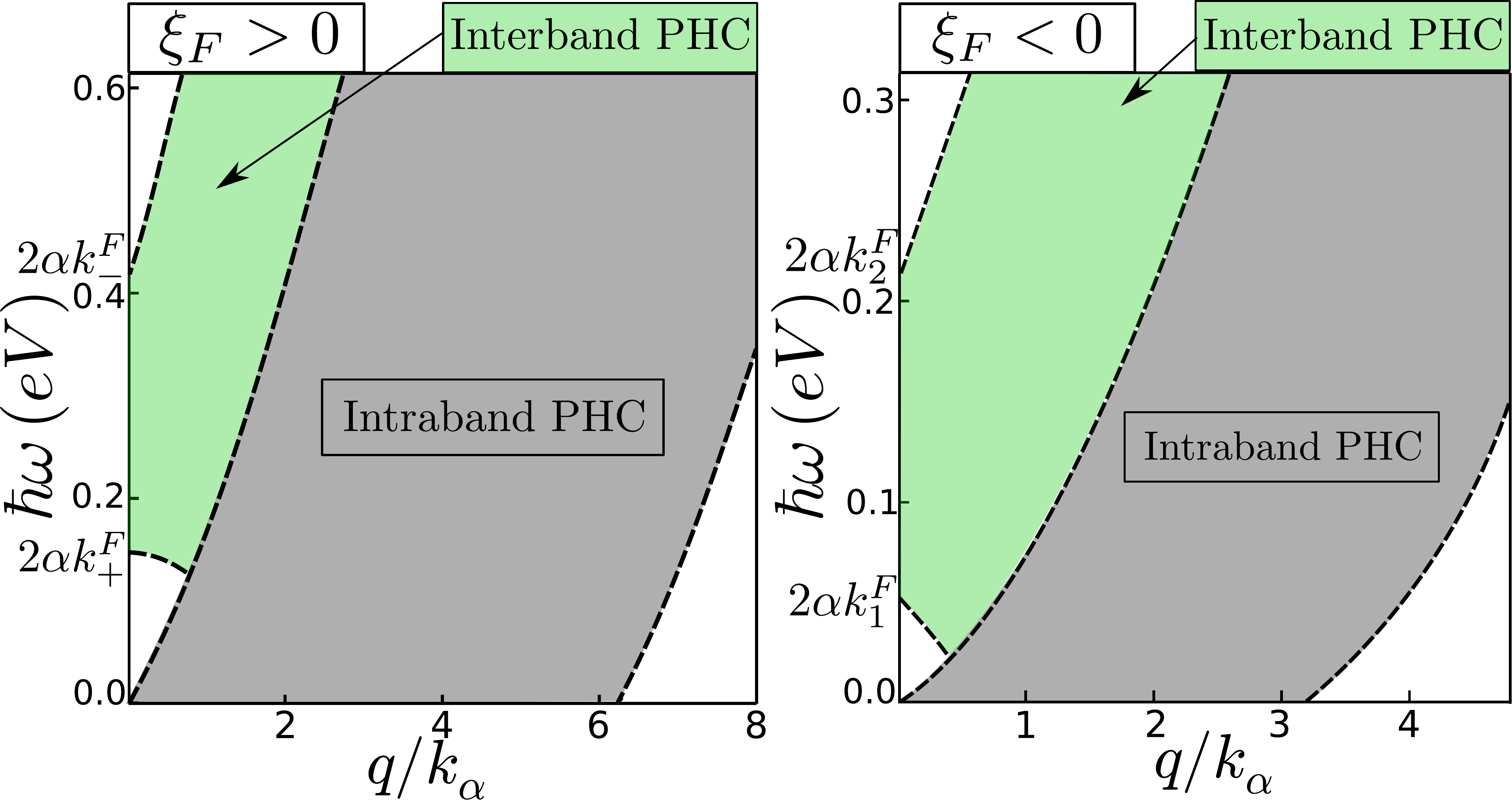}
\caption{Intraband and interband PHC for $\xi_F>0$ (left) and $\xi_F<0$ (right). Minimum and maximum excitation energy for interband  transitions with $q=0$ are $2\alpha k_{+}^F$ ($2\alpha k_{1}^F$) and $2\alpha k_{-}^F$ ($2\alpha k_{2}^F$) for $\xi_F >0$ ($ <0$), respectively. For $\xi_F >0$ ($\xi_F <0$), zero energy intraband transitions ends at $q=2k_{-}^F$ ($q=2k_{2}^F$). It is interesting to note that for all carrier densities ($n_e$) intraband PHC of NCMs is always bigger than that of conventional 3DEG, because $2 k_{F}^0 < 2 k_{-}^F$ ($2 k_{F}^0 < 2 k_{2}^F$) for $\xi_F >0$ ($<0$) with $k_{F}^0=(3\pi^2 n_e)^{2/3}$ being the Fermi wave vector of conventional 3DEG. Parameters: $m^*=0.5 m_0$ with $m_0$ being the bare electron mass, $\alpha = 1~eV\AA$. For left panel $n_e = 16 n_{\alpha}$ and for right panel $n_e = 2 n_{\alpha}$ with $n_{\alpha} =k_{\alpha}^3/(3\pi^2)$. }
\label{PHC}
\end{center}
\end{figure*}
Utilizing the isotropic nature of the band structure, we choose ${\bf q} =q \hat{z}$ for simplicity. 
With $x=k/k_{\alpha}$, $x_{\lambda}^{F}=k_{\lambda}^{F}/k_{\alpha}$, $Q=q/k_{\alpha}$, 
and $D_{\alpha} =m^* k_{\alpha}/(4\pi^2\hbar^2)$, 
performing the $\theta_{\bf k}$ integration exactly, the Lindhard function takes following form for 
$\xi_F>0$ (for $T\rightarrow 0$),
\begin{widetext}
\begin{align}\label{chi0np}
\chi^{0}_{\rho\rho}({\bf q},\Omega)=D_{\alpha}\sum_{\lambda s}\int_{0}^{x_{\lambda}^{F}} 
\frac{dx}{Q} \Big[C_{\lambda}^{s}\log\Big(\frac{t_{\lambda+}^{s}-2Qx}{t_{\lambda+}^{s}+2Qx}\Big)+
G_{\lambda}^{s}\log\Big(\frac{t_{\lambda-}^{s}-2Qx}{t_{\lambda-}^{s}+2Qx}\Big)\Big],
\end{align}
\end{widetext}
with $s=\pm 1$, $\zeta_{\lambda}^{s}=s\hbar\Omega/\xi_{\alpha}+2\lambda x - Q^2$, 
$t_{\lambda\pm}^{s}=s(\zeta_{\lambda}^{s}+2)\pm2 \sqrt{(x+\lambda)^2+s\hbar\Omega/\xi_{\alpha}}$,
$a_{\lambda}^{s}=x(\zeta_{\lambda}^{s_{\omega}}+2\lambda x)$, 
$b_{\lambda}^{s}=s(\lambda-x)$,  
$C_{\lambda}^{s}=(a_{\lambda}^{s} +b_{\lambda}^{s}t_{\lambda+}^{s})/(t_{\lambda+}^{s}-t_{\lambda-}^{s})$, 
and $G_{\lambda}^{s}=-(a_{\lambda}^{s} +b_{\lambda}^{s}t_{\lambda-}^{s})/(t_{\lambda+}^{s}-t_{\lambda-}^{s})$. 
Now it is easy to evaluate this 1D integration numerically with the cost of $s =\pm 1$ summation.
After similar calculation the Lindhard function for $\xi_F<0$ (for $T\rightarrow 0 $) 
\begin{widetext}
\begin{align}\label{chi0nm}
\chi^{0}_{\rho\rho}({\bf q},\Omega)=D_{\alpha}\sum_{s}\int_{x_{1}^{F}}^{x_{2}^{F}} \frac{dx}{Q}\Big[
C_{-}^{s}\log\Big(\frac{t_{-+}^{s}-2Qx}{t_{-+}^{s}+2Qx}\Big)+
G_{-}^{s}\log\Big(\frac{t_{--}^{s}-2Qx}{t_{--}^{s}+2Qx}\Big)\Big],
\end{align}
\end{widetext}
with $s=\pm 1$ and $x_{1}^{F}=k_{1}^{F}/k_{\alpha}$,  $x_{2}^{F}=k_{2}^{F}/k_{\alpha}$. 
Here $k_{\eta}^{F}$ with $\eta=1,~2$ is Fermi wavevector for $\eta$ branch of $\lambda =-$ band 
for $\xi_F<0$. While deriving the above equation, we have used the fact that 
$n_{{\bf k},\lambda}^{F}=0$ for all ${\bf k}$ above the BTP. 
We use Eqs.~\ref{chi0np},~\ref{chi0nm} to present all our numerical results.\\ 
\begin{figure*}[htbp]
	\begin{center}\leavevmode
		\includegraphics[width=0.9\textwidth]{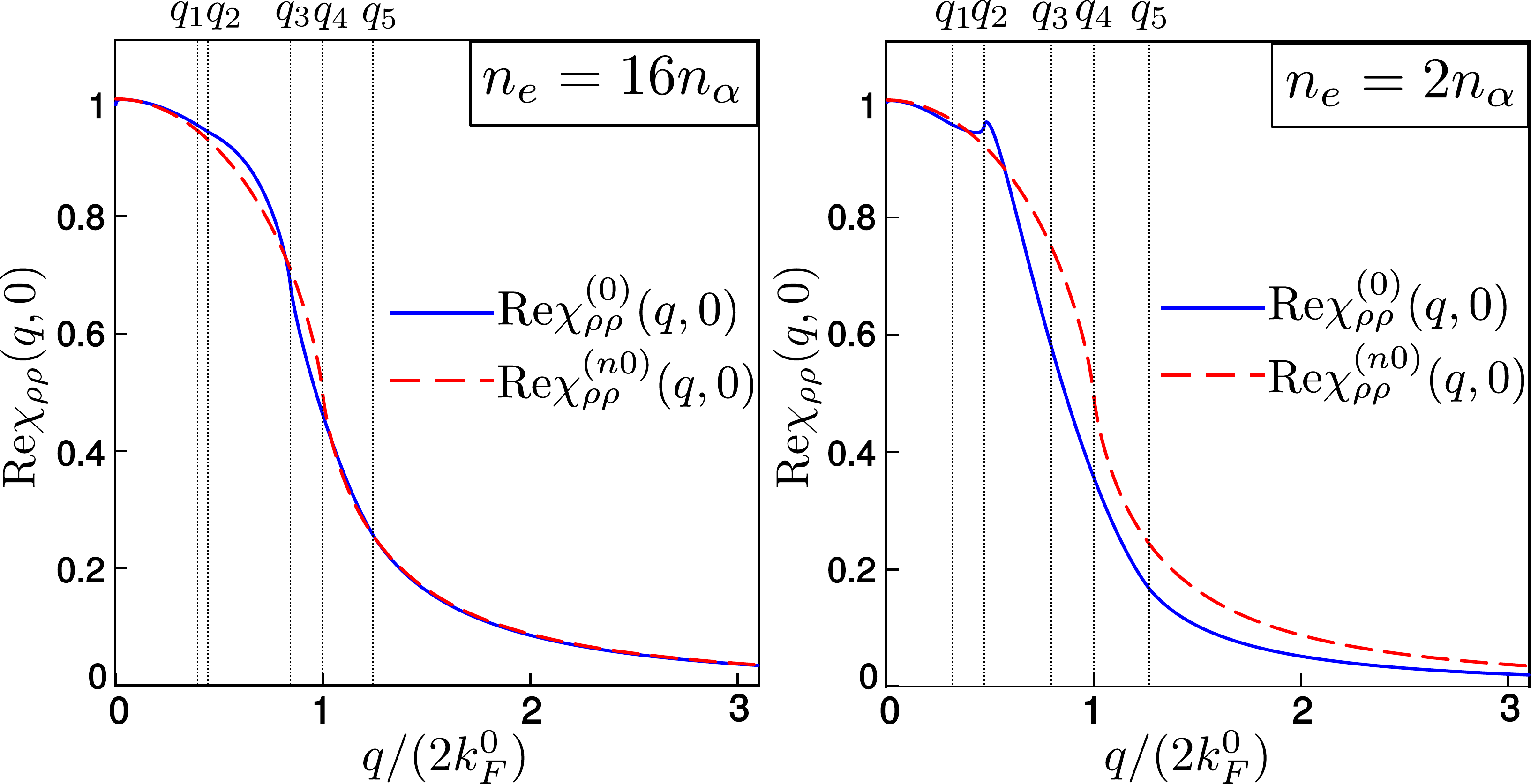}
		\caption{Absolute Static Lindhard function for NCMs ${\rm Re \chi_{\rho\rho}^{0}(q,0)}$ and for conventional 3DEG\cite{giuliani} ${\rm Re \chi_{\rho\rho}^{n0}(q,0)}$ (in units of total density of states of respective systems) vs $q$ for $n_e=16n_{\alpha}$ (left panel) and 
			$n_e=2n_{\alpha}$ (right panel). ${\rm Re \chi_{\rho\rho}^{0}(q,0)}$ is obtained by doing the 1D numerical integration of Eq.~\ref{chi0np} 
			and Eq.~\ref{chi0nm} for $\xi_F>0$ and $\xi_F<0$, respectively. At $n_{e}=2n_{\alpha}$, $\xi_{F}<0$ for NCMs. In this figure, $q_{i}$'s (in units of $2k_{F}^0$) denotes the wavevectors where the Lindhard function or its derivative is singular. In the left panel, these are given as: $q_1 = k_{-}^F - k_{+}^F$, $q_2 = 2k_{+}^F$, $q_3= k_{-}^F + k_{+}^F$, $q_4 = 2k_{F}^0$ and $q_5 = 2k_{-}^F $ in units of $2k_{F}^0$.  And in the right panel, the singular points are: $q_1 = 2k_{1}^F$, $q_2 = k_{2}^F-k_{1}^F$, $q_3= k_{2}^F + k_{1}^F$, $q_4 = 2k_{F}^0$ and $q_5 = 2k_{2}^F $ in units of $2k_{F}^0$.
			Parameters: $m^*=0.5 m_0$, $\alpha = 1~$eV$\AA$. }
		\label{rechi0}
	\end{center}
\end{figure*}
Non-zero ${\rm Im}\chi_{\rho\rho}^{0}({\bf q},\omega)$  for a given (${\bf q}$, $\omega$) describes the 
excitation (with excitation energy $\hbar\omega$) of an electron from an occupied state ${\bf k}$ below 
the Fermi energy to an unoccupied state ${\bf k +q}$ above the Fermi energy and thus leaving a 
hole (empty state) below the Fermi level. The collection of all such points in (${\bf q}$, $\omega$) 
plane is known as particle-hole continuum (PHC). In other words the system can absorb incoming energy by 
exciting 
electron-hole pairs in the region where ${\rm Im}\chi_{\rho\rho}^{0}({\bf q},\omega)\neq 0$. Outside 
the PHC the system can not absorb energy by this mechanism. For NCMs, intra and interband PHC are shown in Fig.~\ref{PHC}. 
The full PHC of NCMs below and above the BTP are of similar nature. The intraband PHC is similar to 
that of noninteracting 3D electron gas. It is worth mentioning here that in NCMs, for $q\rightarrow 0$, there is a finite energy gap in between intra and interband PHC similar to 2D systems with spin-orbit coupling, but it is in contrast to BiTeX semiconductor compounds where the interband 
PHC starts at zero energy.
For $\xi_F>0~(<0)$ the minimum and maximum energy for electron-hole pair excitation with 
$q\rightarrow 0$ is $\hbar\omega = 2\alpha k_{+}^{F}~(2\alpha k_{1}^F)$ and  
$\hbar\omega = 2\alpha k_{-}^F~(2\alpha k_{2}^F)$, respectively.  
The width of interband PHC for $ q\rightarrow 0$ is $\Delta^> =8\xi_{\alpha}$ for $\xi_F>0$ 
and $\Delta^< =8\sqrt{\xi_{\alpha}^2+\xi_{\alpha}\xi_F}$ for $\xi_F < 0$. Due to different Fermi surface 
topology of NCMs for $\xi_F >0$ and $\xi_{F}<0$, $\Delta^>$ and $\Delta^{<}$ show different behaviour 
with respect to the change in the carrier density. Note that $\Delta^{>}$ depends only on the 
Rashba energy $\xi_{\alpha}$, but $\Delta^{<}$ depends on both the carrier density and the Rashba energy\cite{thermNCMs}. This different behavior of width of interband PHC acts as a probe to observe the distinct Fermi surface topology of NCMs for Fermi energies below and above the BTP.\\

Figure~\ref{rechi0} shows the variation of ${\rm Re \chi_{\rho\rho}^0(q,0)}$ with respect to 
the wavevector $q$. Note that 
${\rm Re \chi_{\rho\rho}^0(q,0)}$ is of different nature than that of conventional  3DEG for small $q$ but has similar nature for large $q$. Interestingly, the static Lindhard function of NCMs has distinct second and third derivative singularities owing to the nature 
of distinct Fermi surface topology for $\xi_F>0$ and $\xi_F<0$. The singularities in 
the static Lindhard function arise because of the fact that there 
is a large mismatch of number of states contributing significantly to it below and above the 
singular point. So at the singular point the static Lindhard function changes sharply.
Another way of identifying these singular points is to look for those $q$ for which the original 
Fermi surface $\xi_{{\bf k},\lambda}$ and the shifted Fermi surface $\xi_{{\bf k+q},\lambda^{\prime}}$ 
touch each other. We also show the singular points $q_{i}$ (in units of $2k_{F}^{0}$) in Fig.~\ref{rechi0}. For $\xi_F >0$, the static susceptibility has second derivative singularity at 
$q_3=k_{-}^F+k_{+}^F=2\sqrt{k_{\alpha}^2 + 2m^*\xi_F/\hbar^2}>(k_{-}^F-k_{+}^F)$ due to interband 
transitions similar to the conventional 3DEG and third derivative singularity at $q_2=2k_{+}^F$, $q_5=2k_{-}^F$ arising from the intraband 
transitions. The third derivative singularity at $q_1=k_{-}^F-k_{+}^F=2k_{\alpha}$ is weak. For $\xi_F<0$, the second derivative singularity 
arises at $q_2=k_{2}^F-k_{1}^F=2\sqrt{k_{\alpha}^2 + 2m^*\xi_F/\hbar^2}<(k_{2}^F+k_{1}^F)$ due to 
interbranch transitions and third derivative singularities arise at $q_1=2k_{1}^F$, $q_5=2k_{2}^F$. 
Also the third derivative singularity at $q_3=k_{2}^F+k_{1}^F=2k_{\alpha}$ is weak. Note that the second derivative singularity in the static Lindhard function happens at the addition (difference) of Fermi wavevectors of two bands (branches) for $\xi_F>0(<0)$ as a consequence of change in the Fermi surface topology at the BTP. Although the functional dependence of this singular point on $\alpha$ and $\xi_{F}$ is same for $\xi_F > 0$ and $\xi_F<0$.
The similar nature of singularities in the static Lindhard function was also reported 
in bilayer honeycomb lattice with ultracold atoms\cite{singularbilayer}.\\

\begin{figure*}[ht]
	\centering
	\includegraphics[width=0.9\textwidth]{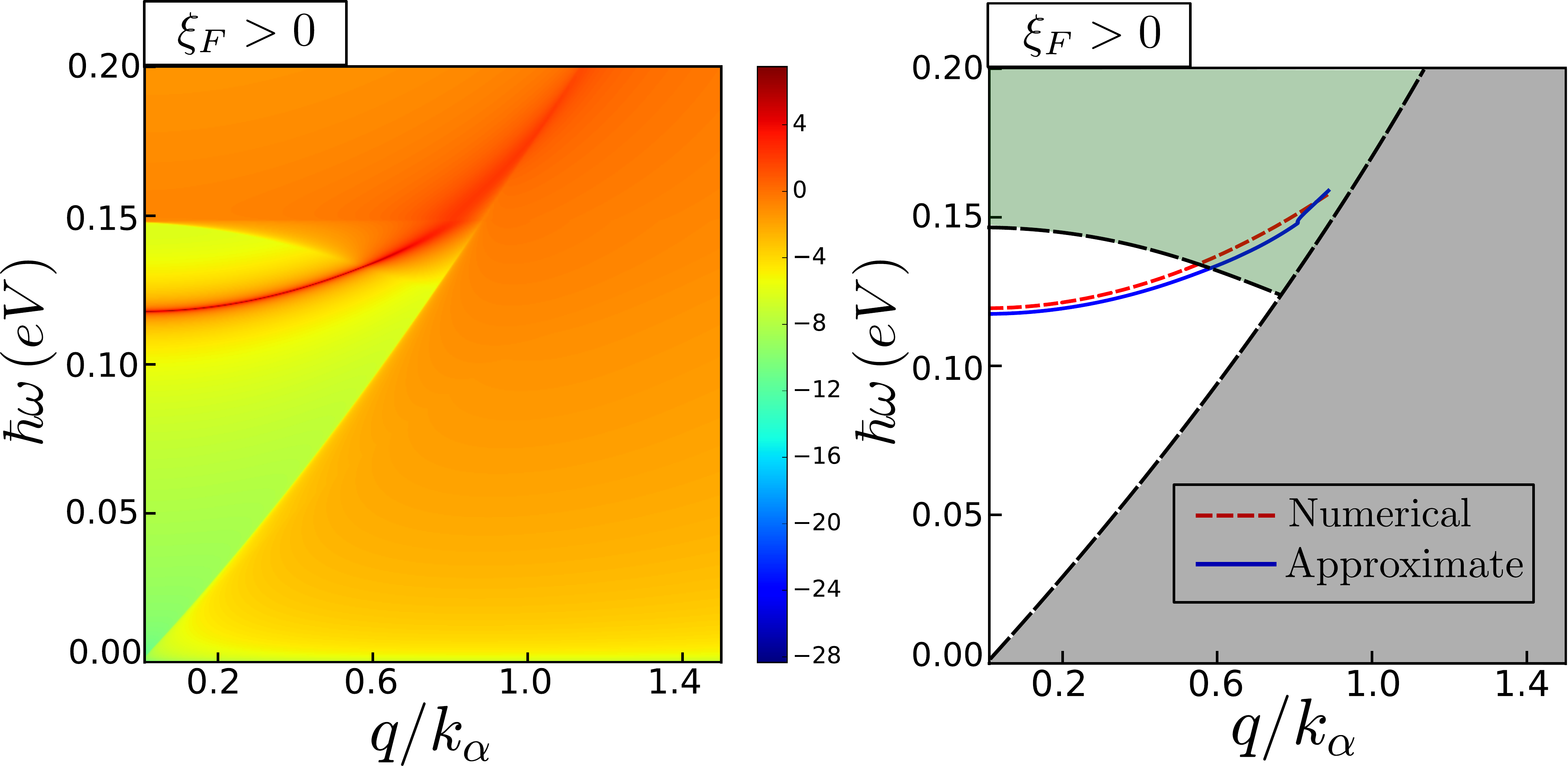}
	\caption{Left panel: Density plot of natural logarithm of loss function defined by Eq.~\ref{lossdefi} for $\xi_F>0$. Sharp bright line shows the undamped plasmon mode outside the PHC. Right panel: Plasmon dispersion together with PHC for $\xi_F>0$. The solid curve shows the plasmon dispersion obtained with the use of Eq.~\ref{epsilon} with exact dynamical polarization function calculated numerically. The dashed curve (apart from the PHC edges) shows the approximate plasmon dispersion given in Eq.~\ref{approxdis}. Parameters: $m^*=0.5m_0$, $\alpha = 1~eV\AA$, $\epsilon_{\infty} =20 \epsilon_{0}$, $n_e =  16 n_{\alpha}$.}
	\label{plasmon_pEF}
\end{figure*}
\begin{figure*}[ht]
	\centering
	\includegraphics[width=0.9\textwidth]{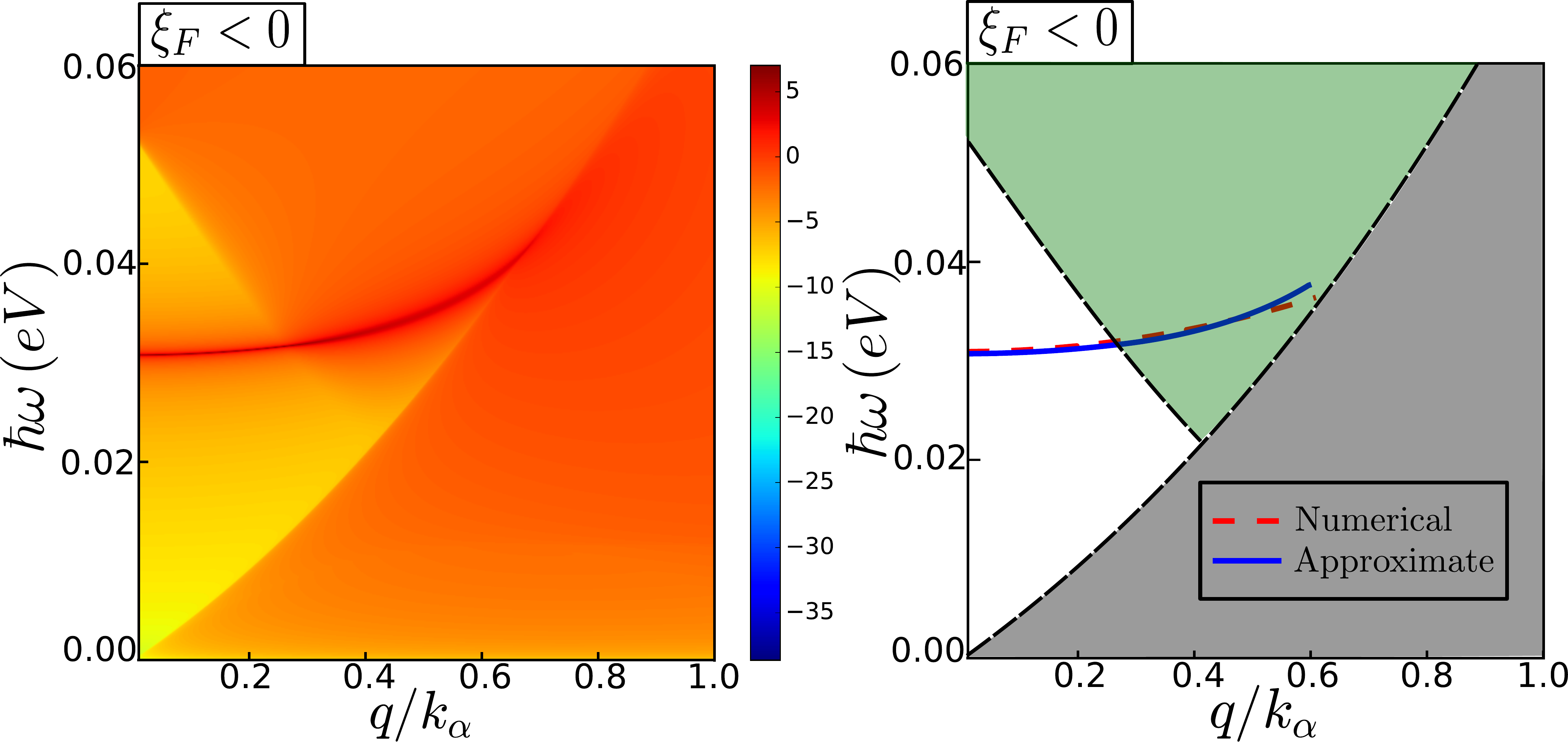}
	\caption{Left panel: Density plot of natural logarithm of loss function defined by Eq.~\ref{lossdefi} for $\xi_F<0$. Sharp bright line shows the undamped plasmon mode outside the PHC. Right panel: Plasmon dispersion together with PHC for $\xi_F<0$. The solid curve shows the plasmon dispersion obtained with the use of Eq.~\ref{epsilon} with exact dynamical polarization function calculated numerically. The dashed curve (apart from the PHC edges) shows the approximate plasmon dispersion given in Eq.~\ref{approxdis}. Parameters: $m^*=0.5m_0$, $\alpha = 1~eV\AA$, $\epsilon_{\infty} =20 \epsilon_{0}$,  $n_e =  2 n_{\alpha}$.}
	\label{plasmon_mEF}
\end{figure*}

\section{Plasmons}
Using the equation of motion technique within RPA the final expression 
of the Lindhard function in presence of the electron-electron interaction 
$\chi^{i}_{\rho\rho}({\bf q},\Omega)$ is given as [see Appendix \ref{app-plasmon}]
\begin{align}\label{chii}
\chi^{i}_{\rho\rho}({\bf q},\Omega) = 
\sum_{\lambda\lambda^{\prime}}\chi_{\lambda\lambda^{\prime}}^{i}({\bf q},\Omega) 
= \frac{\chi_{\rho\rho}^{0}({\bf q},\Omega)}{1-V({\bf q})\chi_{\rho\rho}^{0}({\bf q},\Omega)}.
\end{align} 
Here $\chi^{0}_{\rho\rho}({\bf q},\omega)$ is the dynamical polarization function in the absence of electron-electron interaction which is described in the previous section. The plasmons are described by the poles of the above response 
function $\textit{i.e.}$ zeros of the dielectric function
\begin{align}\label{epsilon}
\epsilon({\bf q},\Omega) = 1-V({\bf q})\chi_{\rho\rho}^{0}({\bf q},\Omega),
\end{align}
with Fourier transform of the Coulomb potential $V({\bf q}) = e^2/(\epsilon_{\infty} q^2)$, 
where $\epsilon_{\infty}=20\epsilon_0$ with $\epsilon_{\infty}$ being the background dielectric constant and $\epsilon_{0}$ is the permittivity of the vacuum. We solve $\epsilon({\bf q},\Omega)=0$ numerically 
using Eq.~\ref{chi0np} and Eq.~\ref{chi0nm} for $\xi_F>0$ and $\xi_F<0$, respectively. 
We first look for plasmon modes for Fermi energy well below and above the BTP. In this case, 
we get two solutions of $\epsilon({\bf q},\Omega)=0$ for a given $q$. Out of these two, the 
higher energy solution lies in between the intra and interband PHC, where 
both ${\rm Re}[\epsilon(q,\omega)] = 0$ 
and ${\rm Im}[\epsilon(q,\omega)] = 0$, which describes the undamped optical plasmon mode. 
Inside intra or interband PHC, ${\rm Im}\chi_{\rho\rho}^{0}({\bf q},\omega)\neq 0$ which 
is responsible for the dissipation in the system. Before reaching the PHC this plasmon mode 
with zero dissipation is an oscillatory eigenmode of the system with infinite life time. 
Inside the PHC this plasmon mode is not an exact eigen mode of the system and acquires 
a finite life time $\propto {\rm Im}\chi_{\rho\rho}^{0}({\bf q},\omega)$. So in this region 
it becomes damped \textit{i.e.} it decays to particle-hole excitations which is also known 
as Landau damping. The other solution fall inside the PHC where ${\rm Im}[\epsilon(q,\omega)] \neq 0$, 
and therefore it is not a solution of $\epsilon({\bf q}, \omega) =0$. The plasmon dispersion together with the PHC for 
a Fermi energy above and below the BTP is shown in the right panels (solid curve) of Fig.~\ref{plasmon_pEF} and Fig.~\ref{plasmon_mEF}, respectively. So here we note that there is only a single undamped optical plasmon mode in NCMs for a range of parameters.  

\begin{figure}[htbp]
	\begin{center}\leavevmode
		\includegraphics[width=0.5\textwidth]{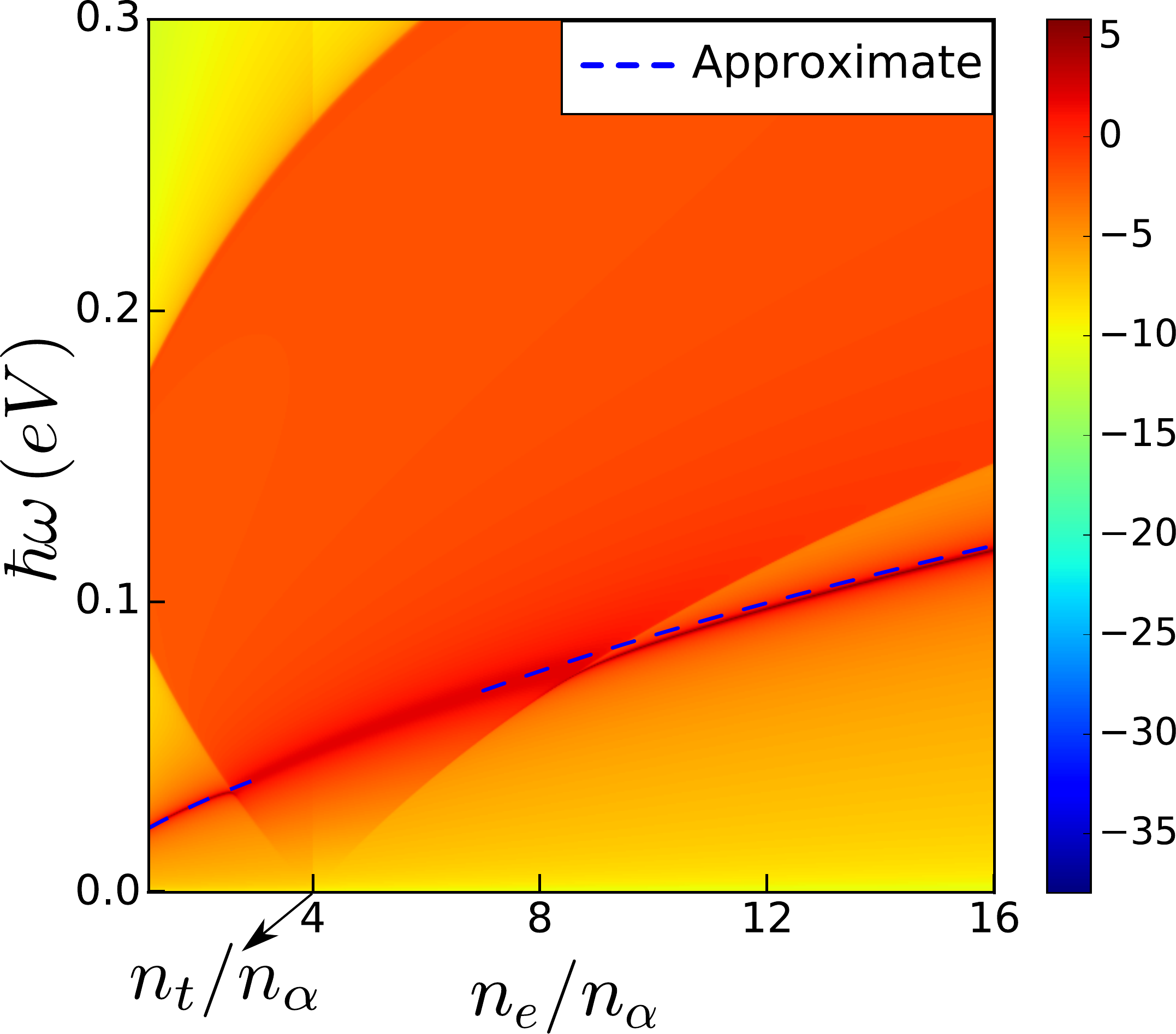}
		\caption{Density plot of natural logarithm of loss function obtained numerically as a function 
			of carrier density $n_e$ (in units of $n_{\alpha}$) and $\hbar\omega$ for small $q$. At carrier density $n_{e}=n_{t}$, $\xi_F=0$ which represents the BTP. The sharp bright line outside the interband PHC indicates that the plasmon mode is undamped only when the Fermi energy $\xi_{F}$ lies well below and above the BTP. Dashed curve shows the approximate plasma frequency obtained from $\omega_{p}^{(>/<)}\approx\omega_{p}^{\prime}/\sqrt{\beta^{(>/<)}}$ (derived in the main text) which matches well with the sharp bright line representing the plasma frequency calculated with the help of exact numerical Lindhard function. Parameters: $m^*=0.5 m_0$, $\alpha = 1~eV\AA$, $\epsilon_{\infty}=20\epsilon_{0}$, and carrier density varies from $n_{e} = 1.1 n_{\alpha}$ to $n_{e} =16.0 n_{\alpha}$. }
		\label{plasmonvsef}
	\end{center}
\end{figure}
\begin{figure*}[ht]
		\begin{center}\leavevmode
			\includegraphics[width=0.9\textwidth]{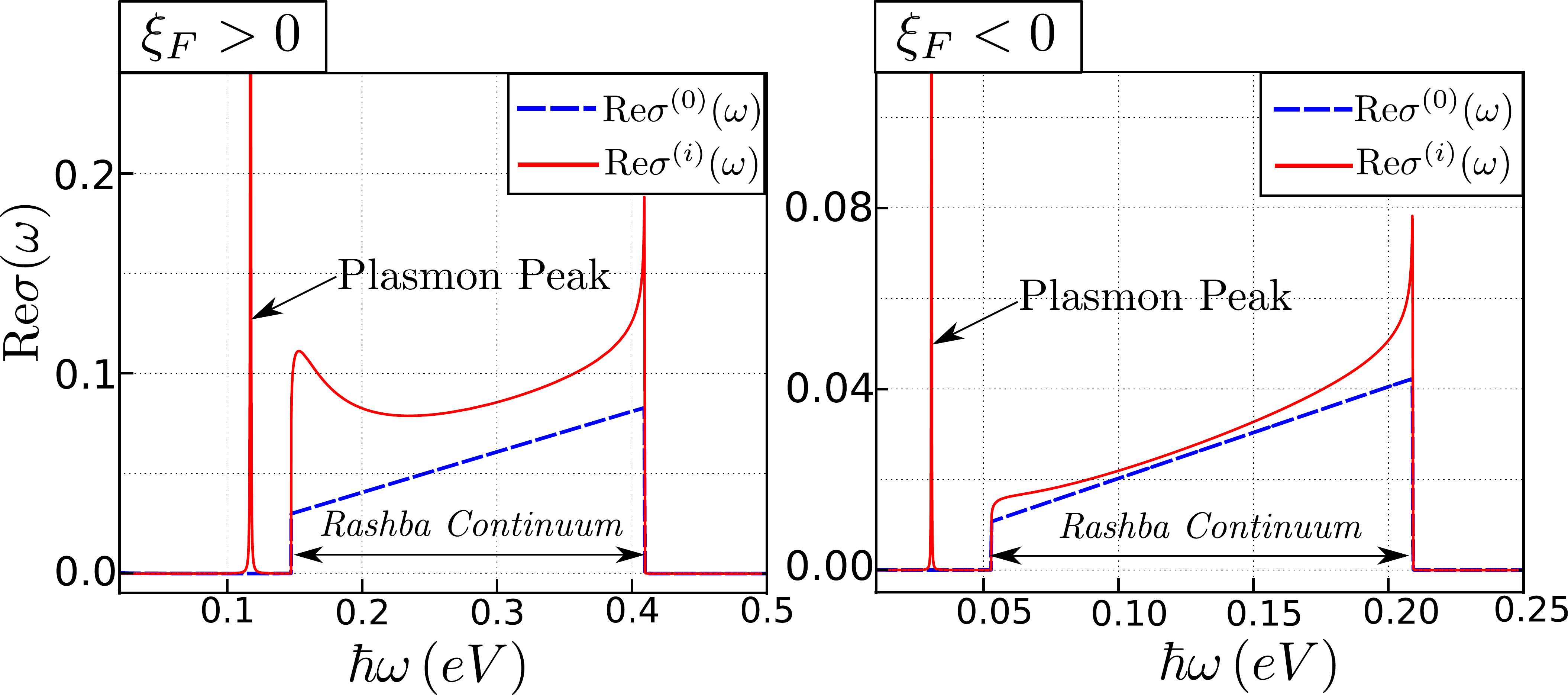}
			\caption{Real part of optical conductivity for $\xi_F>0$ (left panel) 
				and $\xi_F<0$ (right panel). Solid curve shows the behavior of real part of optical conductivity in the presence of electron-electron interaction obtained within RPA. The peak outside the Rasbha continuum indicates the undamped plasmon mode for $q=0.01 k_{\alpha}$. For completeness, we also show the behavior of the real part of optical conductivity for non-interacting case (dashed curve)\cite{thermNCMs}. Parameters: Carrier density $n_e =16 n_{\alpha}$ (same as Fig.~\ref{plasmon_pEF}) for left panel and $n_e =2 n_{\alpha}$ (same as Fig.~\ref{plasmon_mEF}) for left panel, $\epsilon_{\infty}=20\epsilon_{0}$ (same as Fig.~\ref{plasmon_pEF} and Fig.~\ref{plasmon_mEF}). All other parameters are same as Fig.~\ref{PHC}. }
			\label{opt_con}
		\end{center}
\end{figure*}

Now it would be interesting to compare our results with that of in 
BiTeX semiconductor compounds\cite{Opt3}. There are two plasmon modes owing 
to their anisotropic band structure nature in BiTeX semiconductor compounds\cite{Opt3}. 
One out of plane plasmon mode is independent of the in-plane spin-orbit coupling (SOC). 
The other in-plane plasmon mode is dependent on the in-plane SOC but lie within the Rasbha continuum 
and hence it is Landau damped. In these bipolar semiconductor systems the Rashba continuum 
is present for all energies in contrast to 2D Rashba systems\cite{2dplasmon4,2dplasmon2} where it starts at finite energy at $q=0$. 
So the plasmon mode lies within the Rashba continuum for realistic material parameters of 
these systems and hence decays into particle-hole excitations.\\

Plasmon modes can be directly observed in the electron-energy loss and Raman scattering experiments 
by measuring the dynamical structure factor. The dynamical structure factor is proportional 
to the loss function $-{\rm Im}[1/\epsilon(q,\omega)]$. In the left panels of Fig.~\ref{plasmon_pEF} and Fig.~\ref{plasmon_mEF}, we show the density 
plot of the loss function for the Fermi energy well above and below the BTP in ($q$, $\omega$)-plane. 
The loss function can be expressed as
\begin{align}\label{lossdefi}
-{\rm Im}\Big[\frac{1}{\epsilon(q,\omega)}\Big] = \frac{V(q) 
{\rm Im}[\chi_{\rho\rho}^0]}{(1-V(q){\rm Re}[\chi_{\rho\rho}^0])^2+(V(q){\rm Im}[\chi_{\rho\rho}^0])^2}.
\end{align}
From the above expression, it is evident that the loss function is a delta-function for the plasmon 
mode with width of the delta function $\propto {\rm Im}[\chi_{\rho\rho}^0]$. Outside the PHC, 
for the undamped plasmon mode loss function show a well defined delta peak (with very small width due 
to finte $\eta$) which is indicated by a sharp bright line in the left panel of Fig.~\ref{plasmon_pEF} and Fig.~\ref{plasmon_mEF}. As we go inside PHC 
the width of this delta function increases and plasmon mode becomes damped. Also deep inside the PHC, 
the plasmon mode is overdamped and the peak in the loss function disappears which is clearly shown 
in the left panels of Fig.~\ref{plasmon_pEF} and Fig.~\ref{plasmon_mEF}. We note from Fig.~\ref{plasmon_pEF} and Fig.~\ref{plasmon_mEF} that the plasmon dispersion for $\xi_F<0$ is more 
flat than that of $\xi_F>0$. So the plasmon mode has smaller velocity for the Fermi energies below the BTP.\\

In Fig.~\ref{plasmonvsef} for fixed background dielectric constant, the density plot of the loss function in $(\hbar\omega,n_e/n_{\alpha})$ 
plane is shown for small $q$. Sharp bright line shows the behavior of plasma frequency $\omega_p$ 
(defined as the first term in the plasmon dispersion in units of $\hbar$) with respect to 
the carrier density $n_e/n_{\alpha}$ of the system and lighter region compared to the sharp bright line indicates the interband 
PHC for small $q$. For a fixed $\alpha$ at carrier density $n_e = n_{t}$ ($\xi_F=0$) represents the BTP. It is interesting to find that as we tune the Fermi energy around the BTP, the plasmon mode becomes damped within a range of electron-electron interaction strength. Also with a fixed electron-elctron interaction strength when Fermi energy is well below and 
above the BTP, the plasmon mode is undamped, but near the BTP it falls in the interband PHC and 
becomes damped. The reason behind this feature is that the starting point of the Rasbha continuum 
$2\alpha k_{+}^{F}$ ($2\alpha k_{1}^{F}$) for $\xi_{F}>0$ ($<0$) shifts towards 
$\hbar\omega = 0$ as we approach the BTP from above and below. And as a consequence of this, the zero of Eq.~\ref{epsilon} starts to fall within the Rashba continuum.\\ 

Here we provide another known way\cite{Opt3} of observing plasmon modes through optical 
conductivity. It is well known that the finite value of real part of the longitudinal conductivity 
${\rm Re}\sigma ({\bf q},\omega)$ also known as optical conductivity is responsible for the dissipation 
of energy in the system by Joule heating, when a current ${\bf J}({\bf q},\omega)$ is flowing in 
the system. The relation between  ${\rm Im}\chi_{\rho\rho}^{0}({\bf q},\omega)$ and ${\rm Re}\sigma ({\bf q},\omega)$ 
is ${\rm Re}\sigma^{(0)} ({\bf q},\omega) = - \omega e^2 {\rm Im}\chi_{\rho\rho}^{0}({\bf q},\omega)/q^2$. 
So the nonvanishing ${\rm Im}\chi_{\rho\rho}^{0}({\bf q},\omega)$ is also related to the dissipation of the 
energy in the system. From this relation we extract the behavior of real part of optical conductivity ${\rm Re}\sigma^{(0)} ({\bf q},\omega)$ in the absence 
of electron-electron interaction which is shown in Fig.~\ref{opt_con}\cite{thermNCMs}. 
In presence of electron-electron interaction optical conductivity 
becomes (see Appendix \ref{app-op-con}) 
${\rm Re}\sigma^{i} ({\bf q},\omega)=-\omega e^2 {\rm Im}\chi_{\rho\rho}^{i}({\bf q},\omega)/q^2$. 
Here ${\rm Im}\chi_{\rho\rho}^{i}({\bf q},\omega)$ is the dynamical polarization function given 
in Eq.~\ref{chii}. In Fig.~\ref{opt_con}, for Fermi energy above (left panel) and below (right panel) the BTP, we have shown the ${\rm Re}\sigma^{(0)} ({\bf q},\omega)$ and ${\rm Re}\sigma^{(i)} ({\bf q},\omega)$ by dashed and solid lines for small $q$, respectively. The plasmon mode shows up with a peak in ${\rm Re}\sigma^{i} ({\bf q},\omega)$   
between intraband PHC and Rashba continuum as shown in Fig.~\ref{opt_con}.  So from the small $q$ optical conductivity measurement in addition to the measurement of plasma frequency, one can also extract the strength of RSOI ($\alpha$) by measuring the width of the Rashba continuum (same as optical width) which depends differently on carrier density and $\alpha$ for Fermi energies above and below the BTP.\\
 
\begin{figure*}[ht]
	\begin{center}\leavevmode
		\includegraphics[width=0.9\textwidth]{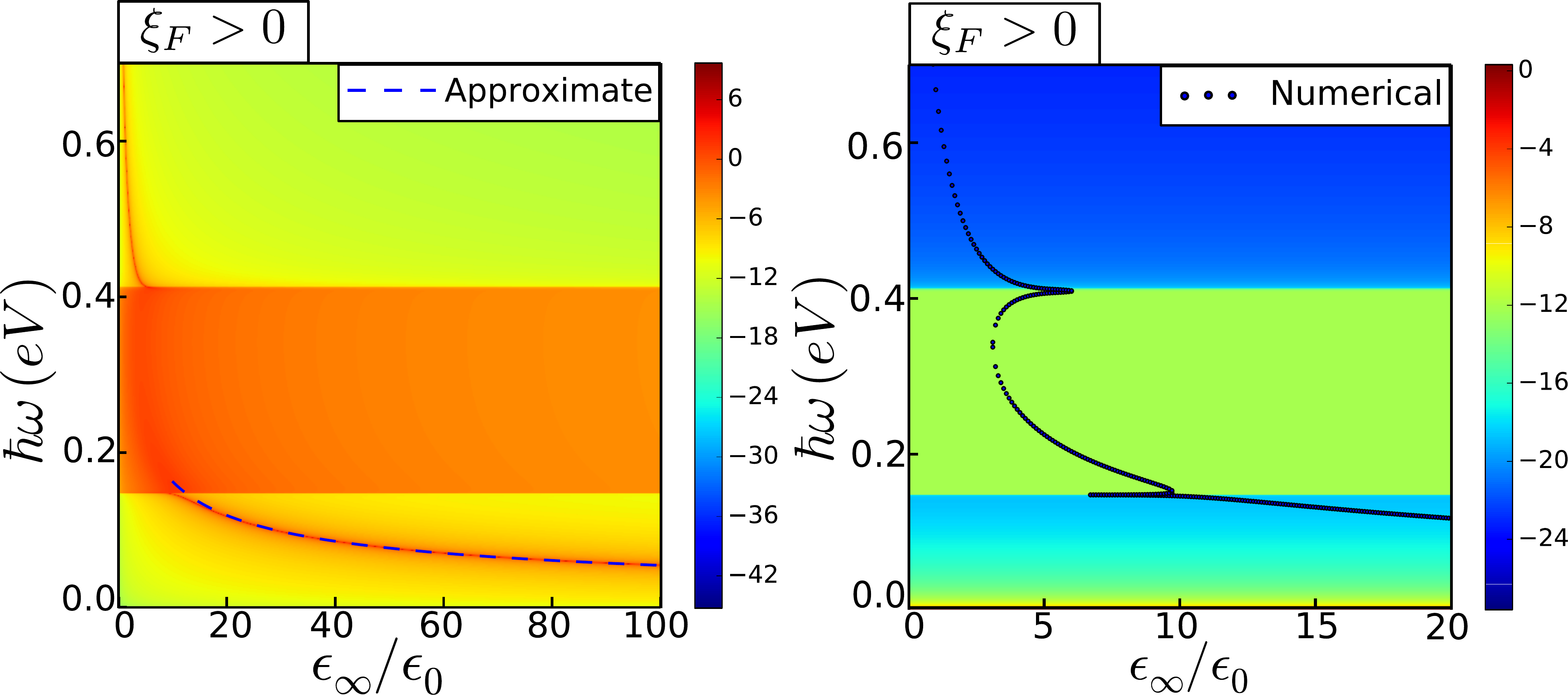}
		\caption{Left panel: Density plot of natural logarithm of loss function in ($\epsilon_{\infty},~\omega$) plane for $\xi_{F}>0$ in the long wavelength limit. Dashed curve shows the variation of approximate plasma frequency $\omega_p$ as a function of $\epsilon_{\infty}$ which determines the interaction strength $\propto 1/\epsilon_{\infty}$.
			Right panel: Density plot of natural logarithm of imaginary part of Lindhard function in  ($\epsilon_{\infty},~\omega$) plane for $\xi_{F}>0$ in the long wavelength limit. Dotted curve shows the plasma frequency as a function of $\epsilon_{\infty}$ obtained numerically from zeros of Eq.~\ref{epsilon} with the help of exact Lindhard function. Parameters: $m^*=0.5m_0$, $\alpha = 1eV\AA$, $n_e = 16 n_{\alpha}$, $q=0.01 k_{\alpha}$.}
		\label{plasmonepsilon_pEF}
	\end{center}
\end{figure*}
\begin{figure*}[ht]
	\begin{center}\leavevmode
		\includegraphics[width=0.9\textwidth]{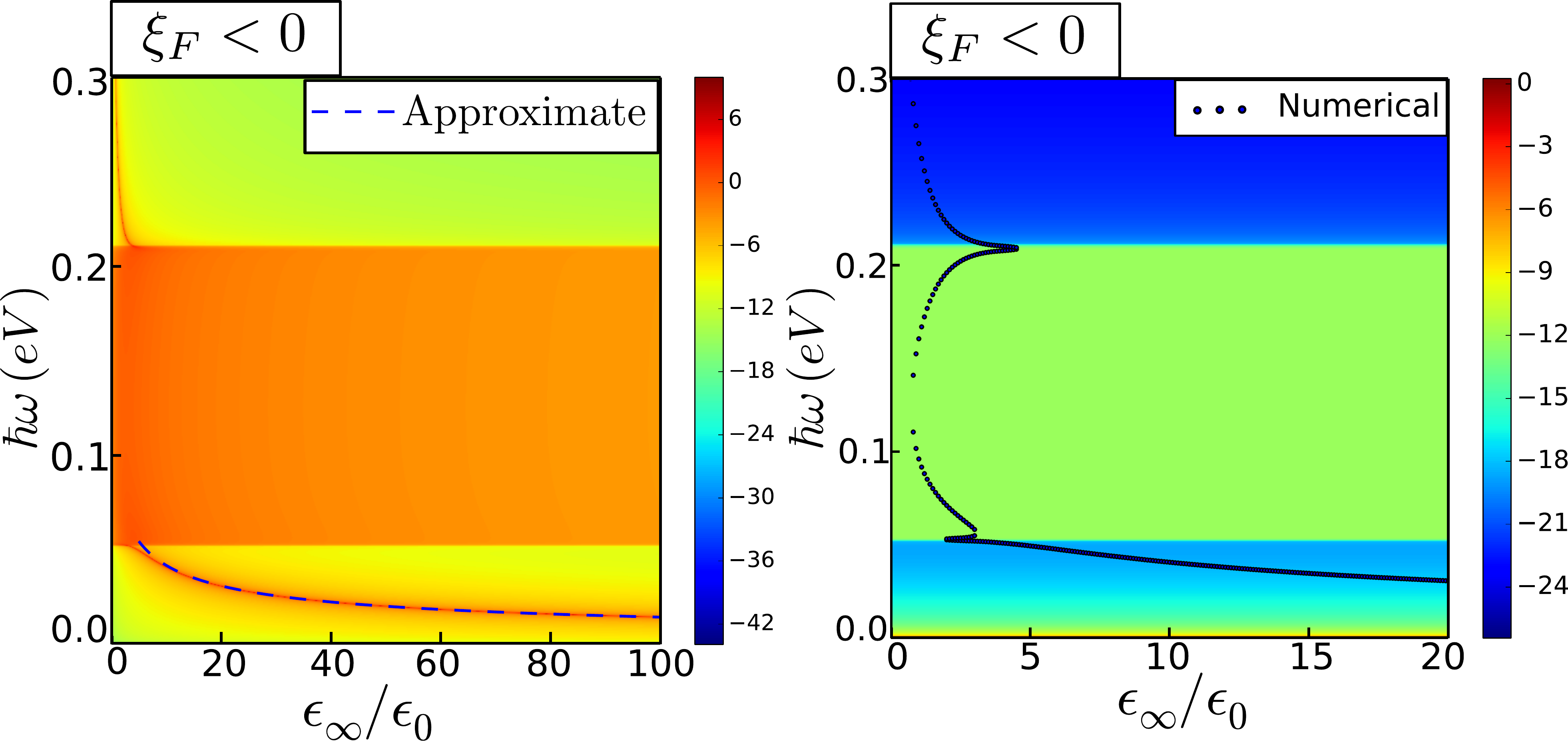}
		\caption{Left panel: Density plot of natural logarithm of loss function in ($\epsilon_{\infty},~\omega$) plane for $\xi_{F}<0$ in the long wavelength limit. Dashed curve shows the variation of approximate plasma frequency $\omega_p$ as a function of $\epsilon_{\infty}$.
			Right panel: Density plot of natural logarithm of imaginary part of Lindhard function in  ($\epsilon_{\infty},~\omega$) plane for $\xi_F <0$ in the long wavelength limit. Dotted curve shows the plasma frequency as a function of $\epsilon_{\infty}$ obtained numerically from zeros of Eq.~\ref{epsilon} with the help of exact Lindhard function. Parameters: $m^*=0.5 m_0$, $\alpha = 1~eV\AA$, $n_e = 2 n_{\alpha}$, $q=0.01 k_{\alpha}$.}
		\label{plasmonepsilon_mEF}
	\end{center}
\end{figure*}
In order to get more insight in the above observations, we derive approximate analytical expressions 
of the plasma frequency $\omega_p$ and plasmon dispersion. For $q \ll k_{\lambda}^{F}$, the full expression 
of $\chi_{\rho\rho}^{0}({\bf q},\omega)$ is given in Appendix \ref{app-approx}. In order to find 
out the plasma frequency 
$\omega_{p}$ we approximate $\chi_{\rho\rho}^{0}({\bf q},\omega)$ for $\xi_F >0$ 
only upto $\mathcal{O}(q^2)$ term, which is given by
\begin{align}\label{approxchi}
\chi_{\rho\rho}^{0}({\bf q},\Omega) & \approx \frac{8D_{\alpha}Q^2\xi_{\alpha}^2[(x_{+}^F)^2+(x_{-}^F)^2] 
\sqrt{1+\xi_F/\xi_{\alpha}}}{3(\hbar\Omega)^2} \nonumber\\
& + \frac{Q^2D_{\alpha}}{6}\log\Big[\frac{(\hbar\Omega)^2-(4\xi_{\alpha}x_{+}^{F})^2}{(\hbar\Omega)^2 - 
(4\xi_{\alpha}x_{-}^{F})^2}\Big].
\end{align}
Here all the notations are the same as in the previous section with $x_{\lambda}^F=k_{\lambda}^F/k_{\alpha}$. 
Then form Eq. (\ref{epsilon}), the plasma frequency $\omega_{p}$ will be given by the zeros of 
the following equation
\begin{align}\label{aprxplsmn}
1-\Big[ \frac{(\hbar\omega_{p}^{\prime})^2}{(\hbar\Omega)^2}+ \frac{D_{\alpha}}{6} 
\log\Big[\frac{(\hbar\Omega)^2-(4\xi_{\alpha}x_{+}^{F})^2}{(\hbar\Omega)^2-(4\xi_{\alpha}x_{-}^{F})^2}\Big]\Big]=0,
\end{align}
with
\begin{align}\label{omprime}
\omega_{p}^{\prime} =\frac{\omega_{p}^{n}}{\sqrt{\epsilon_{\infty}/\epsilon_0}} 
\sqrt{\frac{2\xi_{\alpha}+\xi_F}{4\xi_{\alpha}+\xi_F} },
\end{align}
and  $\omega_{p}^{n}=\sqrt{n_e e^2/m^*\epsilon_0}$ being the plasma frequency for ordinary 3D 
electron gas\cite{giuliani}. The above expression of $\omega_{p}^{\prime}$ has been derived using Eq. (\ref{fermi}). 
We first consider the limiting case when $\alpha =0$. In this case second term in the parenthesis of 
Eq. (\ref{aprxplsmn}) vanishes and putting $\xi_{\alpha} =0$ in Eq. (\ref{omprime}), the plasma 
frequency $\omega_{p}=\omega_{p}^{n}$ reproduces the known  plasma frequency for ordinary 
3D electron gas. In order to achieve an approximate expression of $\omega_p$ for non zero $\alpha$, 
we solve Eq. (\ref{aprxplsmn}) for $\hbar\omega/4\xi_{\alpha}<1$. The plasma frequency becomes 
$\omega_{p}\approx\omega_{p}^{\prime}/\sqrt{\beta^>}$, with 
\begin{align}
\beta^> = 1 - \frac{e^2}{12\pi^2\alpha\epsilon_{\infty}} 
\log\Big[\frac{\sqrt{\xi_{\alpha}+\xi_F}-\sqrt{\xi_{\alpha}}}
{\sqrt{\xi_{\alpha}+\xi_F}+\sqrt{\xi_{\alpha}}}\Big].
\end{align}
We have also obtained similar approximate expression of plasma frequency for Fermi 
energies below the BTP. The plasmon frequency above $(>)$ and below $(<)$ the 
BTP is $\omega_{p}^{(>/<)}\approx\omega_{p}^{\prime}/\sqrt{\beta^{(>/<)}}$, with
\begin{align}
\beta^{(>/<)} = 1-\frac{e^2}{12\pi^2\alpha\epsilon_{\infty}} 
\log\Big[\frac{\pm(\sqrt{\xi_{\alpha}+\xi_F}-\sqrt{\xi_{\alpha}})}{\sqrt{\xi_{\alpha}+\xi_F}+\sqrt{\xi_{\alpha}}}\Big].
\end{align}
The variation of plasma frequency $\omega_p$ with the Fermi energy for fixed background dielectric constant ($\epsilon_{\infty}$) is shown in 
Fig.~\ref{plasmonvsef}. The sharp bright line describes numerically obtained $\omega_p$ and 
on top of that dashed lines describes the analytical result obtained from 
$\omega_{p}^{(>/<)}\approx\omega_{p}^{\prime}/\sqrt{\beta^{(>/<)}}$. \\ 

We also find an approximate analytical expression of plasmon dispersion with the help of 
the approximate plasma frequency $\omega_{p}^{(>/<)}$ and $\chi_{\rho\rho}^{0}({\bf q}, \omega)$ 
in the long wavelength limit. The plasmon dispersion in the long wavelength limit above and below BTP becomes
\begin{align}\label{approxdis}
\omega^{(>/<)}(q)&\approx \omega_{p}^{(>/<)} + \frac{3}{5}\frac{(\xi_{\alpha}+\xi_F)}{m^*\omega_{p}^{(>/<)}}q^2\nonumber\\
&\approx \omega_{p}^{(>/<)} + \frac{3}{10}\frac{(v_{>/<}^{F}q)^2}{\omega_{p}^{(>/<)}},
\end{align}
with $v_{>}^{F}$ ($v_{<}^{F}$) is the absolute value of the Fermi velocity for $\xi_{F}>0$ ($<0$) which can be expressed as $v_{>}^{F}=\hbar (k_{\lambda}^{F}+\lambda)/m^*=(\hbar k_{\alpha}/m^*)\sqrt{1+\xi_F/\xi_{\alpha}}$ and $v_{<}^{F}=\hbar |(k_{\eta}^{F}-1)|/m^*=(\hbar k_{\alpha}/m^*)\sqrt{1+\xi_F/\xi_{\alpha}}$. The absolute value of Fermi velocity increases with increase of carrier density or equivalentally Fermi energy. This implies that the plasmon mode for Fermi energy below the BTP has smaller velocity ($\propto \sqrt{\xi_F + \xi_{\alpha}}$) than the plasmon mode for Fermi energy above the BTP as mentioned earlier from numerical results (see right panels of Figs.~\ref{plasmon_pEF},~\ref{plasmon_mEF}). Also for all carrier densities, the stronger spin-orbit coupling reduces the Fermi energy and Fermi velocity ($\propto \sqrt{\xi_F + \xi_{\alpha}}$) of the system, so the plasmon velocity ($\propto \sqrt{\xi_F + \xi_{\alpha}}$) also decreases. The above equation also indicates that in the long wavelength limit the plasmon dispersion  is $\propto q^2$ which is similar to that of 
ordinary 3D electron gas\cite{giuliani}. This approximate plasmon dispersion has been shown in the right panels of Figs.~\ref{plasmon_pEF},~\ref{plasmon_mEF} together with the exact plasmon dispersion obtained numerically. It is evident from right panels of Figs.~\ref{plasmon_pEF},~\ref{plasmon_mEF} and Fig.~\ref{plasmonvsef} that the approximate plasma frequency and plasmon dispersion matches well with the exact numerical dispersion in the long wavelength limit when the excitation energy for the plasmon are smaller than $4\xi_{\alpha}$. For higher or comparable excitation energy which happens at larger carrier densities, the approximate plasmon dispersion starts to deviate from the exact numerical dispersion.
Also as we have already discussed that in the limiting case \textit{i.e.} 
$\alpha =0$, $\omega_p = \omega_{p}^{n}$. 
Applying these to above equation for plasmon dispersion reproduces the correct form of the plasmon 
dispersion in the long-wavelength limit for ordinary 3D electron gas\cite{giuliani}. 


For all the results from Fig.~\ref{plasmon_pEF} to Fig.~\ref{opt_con} shown above we have taken the background dielectric constant\cite{Opt3,epsilonBiTeI} $\epsilon_{\infty} = 20\epsilon_0$. We  also show in Figs.~\ref{plasmonepsilon_pEF}, ~\ref{plasmonepsilon_mEF} that for fixed carrier densities above and below the BTP, changing the strength of electron-electron interaction which is inversely proportional to the background dielectric constant $\epsilon_{\infty}$ for the Fermi energy above and below the BTP does not change the number of undamped plasmon modes, although the damped plasmon modes in the interband PHC are more in number. It is also clear from Fig.~\ref{plasmonepsilon_pEF} and Fig.~\ref{plasmonepsilon_mEF} that for small $\epsilon_{\infty}$ the plasma frequency decreases rapidally and after that decreases slowly with further increase in $\epsilon_{\infty}$. As the Fermi energy is fixed, the interband PHC is also fixed  and only the zeros of $\epsilon({\bf q}, \omega) =0$ are changing with $\epsilon_{\infty}$. In the left panel of Figs.~\ref{plasmonepsilon_pEF},~\ref{plasmonepsilon_mEF}, dashed line shows the variation of approximate plasma frequency $\omega_{p}$ with respect to $\epsilon_{\infty}$. The approximate plasma frequency matches well with it's numerical counterpart for larger background dielectric constant.

\section{Summary and Discussion}
In summary, we have studied the dynamical polarization function and plasmon modes of NCMs in detail. In NCMs, the Rashba continuum is similar to that of 2DEG with spin-orbit coupling, and it starts at finite energy in contrast to the BiTeX semiconductor compounds\cite{Opt3}. In the long wavelength limit, the width of Rashba continuum behaves differently for Fermi energies below and above the BTP as a consequence of change in the Fermi surface topology. Within a range of electron-electron interaction strength and of suitable material parameters, there is a single undamped optical plasmon mode for Fermi energies above and below the BTP. Interestingly we find that the plasmon mode is damped for Fermi energies near the BTP within a range of electron-electron interaction strength. For fixed carrier densities above and below the BTP, with the increase of background dielectric constant, the number of undamped plasmon mode does not change, although the damped plasmon modes can be more in number.  It is important to note here that for a fixed electron-electron interaction strength and a range of Fermi energies or vice-versa with other material parameters NCMs always has one undamped plasmon mode. So for the same range of realistic material parameters, NCMs host a single undamped plasmon mode whereas the plasmon modes are always damped in BiTeX semiconductor compounds\cite{Opt3}. \\
In NCMs, the approximate plasma frequency and the plasmon dispersion ($\propto q^2$) matches well with the exact numerical results in the long wavelength limit. The velocity of plasmon mode is $\propto\sqrt{\xi_F + \xi_{\alpha}}$. So for Fermi energies below the BTP, plasmon mode has smaller velocity compared to that of Fermi energies above the BTP.
 At fixed electron-electron interaction strength, the plasma frequency has similar carrier density dependence for Fermi energies above and below the BTP.  For Fermi energies above and below the BTP, the plasma frequency decreases rapidally for smaller $\epsilon_{\infty}$ and after that decreases slowly with further increase in $\epsilon_{\infty}$. The approximate plasma frequency as a function of $\epsilon_{\infty}$ also matches well with the exact numerical result for larger $\epsilon_{\infty}$. It is important to note that the approximate analytical expression of plasma frequency and plasmon dispersion are valid for $\hbar\omega/4\xi_{\alpha}<1$ in the long wavelength limit. \\
It should be mentioned here that for small $\beta$ the presence of cubic spin-orbit coupling term in the effective Hamiltonian may not change the dielectric properties significantly. However for large $\beta$, it may give rise to anisotropic plasmon modes similar to the 2D electron\cite{2dplasmon4} and hole gas\cite{2DHGanisotropic}. Moreover, if the bands around other symmetry points in Brillouin zone cross the Fermi energy, they may also contribute to the dielectric properties in some form\cite{B201}. \\
\begin{center}
{\bf ACKNOWLEDGEMENTS}
\end{center}
We would like to thank H. A. Fertig (IU Bloomington) for some useful discussion. A.K. and S.V. acknowledge support from the SERB (Govt. of India) via sanction no. ECR/2018/001443, DAE (Govt. of India ) via sanction no. 58/20/15/2019-BRNS, as well as MHRD (Govt. of India) via sanction no. SPARC/2018-2019/P538/SL. We also acknowledge HPC facility of IIT Kanpur for computational work.

\appendix{}
\begin{widetext}

\section{The second quantized representation}\label{ham-2ndquan} \label{app-pf}
In order to study many body systems, it is convenient to work in the occupation number or second 
quantized representation\cite{giuliani,bruusflensberg}. For translationally invariant systems, we choose a single particle basis 
$\lbrace |\textbf{k} \sigma\rangle\rbrace$ with $\sigma=\uparrow,\downarrow$ and 
$\langle \textbf{r}|\textbf{k} \sigma\rangle = \tilde{\psi}_{\textbf{k},\sigma}(\textbf{r})=
\eta_{\sigma} e^{i {\bf k \cdot r}}/\sqrt{\mathcal{V}}$ with $\eta_{\uparrow} = 
\lbrace1~0\rbrace^{\mathcal{T}}$ and $\eta_{\downarrow} = \lbrace0~1\rbrace^{\mathcal{T}}$, 
$\mathcal{T}$ stands for transpose. As particles are indistinguishable, the basis states in 
the occupation number representation is $\lbrace|n_{\textbf{k} \sigma}\rangle\rbrace$ such that 
$\sum_{\textbf{k}\sigma} n_{\textbf{k}\sigma}=N$, where $N$ is the total number of particles.  
We define electron creation $\tilde{C}^{\dagger}_{\textbf{k},\sigma}$ and annihilation operator 
$\tilde{C}_{\textbf{k},\sigma}$  with spin $\sigma$ which increases and decreases the occupation 
number of state $|n_{\textbf{k} \sigma}\rangle$ by unity, respectively. All first quantized operators 
can be expressed in the second quantized form using the quantum field operators defined as 
\begin{align}
\tilde{\Psi}^{\dagger}(\textbf{r})=\sum_{\textbf{k},\sigma}\frac{e^{-i {\bf k \cdot r}}}{\sqrt{\mathcal{V}}}
\eta^{\dagger}_{\sigma} \tilde{C}^{\dagger}_{{\bf k},\sigma}
~~{\rm and}~~\tilde{\Psi}(\textbf{r})=\sum_{\textbf{k},\sigma}\frac{e^{i {\bf k \cdot r}}}{\sqrt{\mathcal{V}}}
\eta_{\sigma} \tilde{C}_{{\bf k},\sigma}.
\end{align} 
Density operator in second quantized form is given by
\begin{align}
\hat{\rho}(\textbf{r}) & = \int d\textbf{r}^{\prime}\tilde{\Psi}^{\dagger}(\textbf{r}^{\prime}) 
\delta(\textbf{r}-\textbf{r}^{\prime}) \tilde{\Psi}(\textbf{r}^{\prime}), \nonumber\\
& = \tilde{\Psi}^{\dagger}(\textbf{r})\tilde{\Psi}(\textbf{r}),\nonumber\\
& = \frac{1}{\mathcal{V}}\sum_{{\bf q}}e^{i {\bf q}\cdot{\bf r}}\hat{\rho}({\bf q}),
~~{\rm with}~~ \hat{\rho}({\bf q}) = 
\sum_{{\bf k}\sigma}\tilde{C}^{\dagger}_{\textbf{k},\sigma}\tilde{C}_{\textbf{k+q},\sigma}.
\end{align}
The Hamiltonian is diagonal in the helicity basis ${|{\bf k}\lambda\rangle}$ with $\lambda =\pm$. 
We define quantum field operators in this basis as
\begin{align}
\Psi^{\dagger}(\textbf{r})=\sum_{\textbf{k},\lambda}\frac{e^{-i {\bf k \cdot r}}}{\sqrt{\mathcal{V}}}
\phi^{\dagger}_{{\bf k},\lambda} C^{\dagger}_{{\bf k},\lambda}
~~{\rm and}~~\Psi(\textbf{r})=\sum_{\textbf{k},\lambda}\frac{e^{i {\bf k \cdot r}}}{\sqrt{\mathcal{V}}}
\phi_{{\bf k},\lambda} C_{{\bf k},\lambda}.
\end{align} 

Now the Hamiltonian $H_0$ and the density operator in the second quantized form in 
the helicity basis are
\begin{align}
\hat{H}_0 & = \sum_{{\bf k},\lambda}\xi_{{\bf k},\lambda}
C^{\dagger}_{\textbf{k},\lambda} C_{\textbf{k},\lambda},
~~{\rm with}~~\xi_{{\bf k},\lambda}=\hbar^2k^2/(2 m^*) + \lambda \alpha k. \nonumber\\
\hat{\rho}({\bf q}) & = \sum_{{\bf k}\lambda_1\lambda_2}\phi^{\dagger}_{{\bf k},\lambda_1} 
\phi_{{\bf k +q},\lambda_2}C^{\dagger}_{\textbf{k},\lambda_1}C_{\textbf{k+q},\lambda_2}.
\end{align}

Consider the perturbed Hamiltonian 
$\hat{H}(t) = \hat{H}_{0} + \int d{\bf r} V_{\rm ext}({\bf r},t)\hat{\rho}({\bf r})$. 
The induced density due to this perturbation is given by\cite{giuliani,bruusflensberg}
\begin{align}
\rho_{\rm ind}({\bf r},t) = \int_{-\infty}^{t} d t^{\prime} \int d{\bf r}^{\prime} 
\chi^{0}_{\rho\rho}({\bf r},{\bf r}^{\prime},t,t^{\prime}) V_{\rm ext}({\bf r}^{\prime},t^{\prime}).
\end{align} 
Here $\chi^{0}_{\rho\rho}({\bf r},{\bf r}^{\prime},t,t^{\prime})$ which is known as 
the retarded density-density 
response function, is the response of the density operator averaged over the ground state of perturbed 
Hamiltonian due to the perturbation. The $`\rho\rho'$ in the subscript indicates that it is density-density 
correlation function. The induced density is $\rho_{\rm ind}({\bf r},t) 
\equiv \langle \hat{\rho}({\bf r},t)\rangle_{{\rm ext}}-\langle \hat{\rho}({\bf r},t)\rangle_{0}$. 
The symbols $\langle ... \rangle_{\rm ext}$ and $\langle ... \rangle_{0}$ denotes the average is taken 
over the ground state of the perturbed $\hat{H}(t)$ and unperturbed Hamiltonian $\hat{H}_0$. 
Within the linear response formalism\cite{giuliani,bruusflensberg}, the retarded density-density response function has following form
\begin{align}
\chi^{0}_{\rho\rho}({\bf r},{\bf r}^{\prime},t,t^{\prime})=-\frac{i}{\hbar}\theta(t-t^{\prime}) 
\langle[\hat{\rho}({\bf r},t),\hat{\rho}({\bf r}^{\prime},t^{\prime})]\rangle_{0}.
\end{align} 
For translationally invariant systems, the density-density response function in Fourier space 
is given by
\begin{align}
\chi^{0}_{\rho\rho}({\bf q},t,t^{\prime}) = - \frac{i}{\hbar\mathcal{V}}\theta(t-t^{\prime}) 
\langle[\hat{\rho}({\bf q},t),\hat{\rho}(-{\bf q},t^{\prime})]\rangle_{0}.
\end{align} 
In above expressions the time dependence of the operators comes in the form 
$\hat{A}(t) = e^{i \hat{H}_{0}t/\hbar}\hat{A}(0)e^{-i \hat{H}_{0}t/\hbar}$. 
After some straight forward algebra the final expression of the density-density response function in 
Fourier space $\chi^{0}_{\rho\rho}({\bf q},\omega)=\int_{-\infty}^{+\infty}d t~e^{i\omega(t-t^\prime)} 
\chi^{0}_{\rho\rho}({\bf q},t-t^{\prime})$ becomes
\begin{align}\label{chi0A}
\chi^{0}_{\rho\rho}({\bf q},\omega)=\sum_{\lambda\lambda^{\prime}} 
\chi^{0}_{\lambda\lambda^{\prime}}({\bf q},\omega),
~~{\rm with}~~ \chi^{0}_{\lambda\lambda^{\prime}}({\bf q},\omega) 
= 
\frac{1}{\mathcal{V}}\sum_{\bf k}\mathcal{F}_{\lambda\lambda^{\prime}}({\bf k},{\bf k+q}) 
\frac{n^{F}_{{\bf k},\lambda} - n^{F}_{{\bf k+q},\lambda^{\prime}} }
{\hbar(\omega+i 0^{+})+\xi_{{\bf k},\lambda}-\xi_{{\bf k+q},\lambda^{\prime}}}.
\end{align}
Here $\mathcal{F}_{\lambda\lambda^{\prime}}({\bf k},{\bf k+q}) 
= |\phi^{\dagger}_{{\bf k},\lambda}\phi_{{\bf k +q},\lambda^{\prime}} |^2$ describes the 
overlap between the two states labelled by $|{\bf k},\lambda\rangle$ 
and $|{\bf k+q},\lambda^{\prime}\rangle$. 
Also, $n^{F}_{{\bf k},\lambda} = 1/[e^{-\beta(\xi_{{\bf k},\lambda}-\mu)} + 1]$ is the
Fermi-Dirac distribution function with $\beta=(k_{\rm B} T)^{-1}$ and $T$ being the temperature.\\ 

In order to get the full ${\bf q}$ and $\omega$ dependence of $\chi_{\rho\rho}^{0}({\bf q},\omega)$ 
first we simplify its expression for appropriate numerical simulation. We have also derived an 
asymptotic expression of $\chi_{\rho\rho}^{0}({\bf q},\omega)$ for $q \ll k^{F}_{\lambda/\eta}$ which 
we will describe in the later section.
Using the ground state properties of NCS metals, we simplify $\chi^{0}_{\rho\rho}({\bf q},\omega)$ 
for $\xi_F > 0$ as follows
\begin{align}
\chi^{0}_{\rho\rho}({\bf q},\omega) & =\frac{1}{2\mathcal{V}}\sum_{{\bf k}\lambda\lambda^{\prime}} 
\Big[1 + \lambda\lambda^{\prime}\frac{{\bf k}\cdot({\bf k+q})}{|{\bf k}| |{\bf k+q}|} \Big] 
\frac{n^{F}_{{\bf k},\lambda} - n^{F}_{{\bf k+q},\lambda^{\prime}} }
{\hbar\Omega+\xi_{{\bf k},\lambda}-\xi_{{\bf k+q},\lambda^{\prime}}},\nonumber\\
&=\chi^{0(+)}_{\rho\rho}({\bf q},\omega)+\chi^{0(-)}_{\rho\rho}({\bf q},\omega),
\end{align}
where $\chi^{0(+)}_{\rho\rho}({\bf q},\omega)$ has the following expression (for $T\rightarrow 0$)
\begin{align}
\chi^{0(+)}_{\rho\rho}({\bf q},\omega) & = \frac{1}{2\mathcal{V}}\sum_{{\bf k}\lambda\lambda^{\prime}} 
\Big[1 + \lambda\lambda^{\prime}\frac{{\bf k}\cdot{\bf (k+q)}}{|{\bf k}| |{\bf k+q}|} \Big] 
\frac{n^{F}_{{\bf k},\lambda}}{\hbar\Omega+\xi_{{\bf k},\lambda}-\xi_{{\bf k+q},\lambda^{\prime}}},\nonumber\\
& = D_{\alpha}\sum_{\lambda}\int_{0}^{x_{\lambda}^{F}}x^2 dx\int_{0}^{\pi}\sin\theta_{\bf k} d\theta_{\bf k}
\Big[ \frac{2(\zeta_{\lambda}^{+}-2Qx\cos\theta_{\bf k}) + 
4\lambda(x+Q\cos\theta_{\bf k})}{(\zeta_{\lambda}^{+}-2Qx\cos\theta_{\bf k})^2-4|{\bf x + Q }|^2}\Big],
\end{align}
with $x=k/k_{\alpha}$, $x_{\lambda}^{F}=k_{\lambda}^{F}/k_{\alpha}$, $Q=q/k_{\alpha}$, 
$D_{\alpha} =m^* k_{\alpha}/(4\pi^2\hbar^2)$, and 
$\zeta_{\lambda}^{+}=\hbar\Omega/\xi_{\alpha}+2\lambda x - Q^2$. After doing the straight forward 
$\theta_{\bf k}$ integration, $\chi^{0(+)}_{\rho\rho}({\bf q},\omega)$ has following form
\begin{align}
\chi^{0(+)}_{\rho\rho}({\bf q},\omega)=D_{\alpha}\sum_{\lambda}\int_{0}^{x_{\lambda}^{F}} \frac{dx}{Q}
\Big[ C_{\lambda}^{+}\log\Big(\frac{t_{\lambda+}^{+}-2Qx}{t_{\lambda+}^{+}+2Qx}\Big)+
G_{\lambda}^{+}\log\Big(\frac{t_{\lambda-}^{+}-2Qx}{t_{\lambda-}^{+}+2Qx}\Big)\Big],
\end{align}
with $t_{\lambda\pm}^{+}=(\zeta_{\lambda}^{+}+2)\pm 2\sqrt{(x+\lambda)^2+\hbar\Omega/\xi_{\alpha}}$, 
$C_{\lambda}^{+}=(a_{\lambda}^{+} +b_{\lambda}^{+}t_{\lambda+}^{+})/(t_{\lambda+}^{+}-t_{\lambda-}^{+})$, 
$G_{\lambda}^{+}=-(a_{\lambda}^{+} +b_{\lambda}^{+}t_{\lambda-}^{+})/(t_{\lambda+}^{+}-t_{\lambda-}^{+})$, 
$a_{\lambda}^{+}=x(\zeta_{\lambda}^{+}+2\lambda x)$, and $b_{\lambda}^{+}=\lambda-x$. 
The above 1D integration can be done numerically. Now let's consider $\chi^{0(-)}_{\rho\rho}({\bf q},\omega)$
\begin{align}
\chi^{0(-)}_{\rho\rho}({\bf q},\omega)&=\frac{1}{2\mathcal{V}}\sum_{{\bf k}\lambda\lambda^{\prime}}
\Big[1+\lambda\lambda^{\prime}\frac{{\bf k}\cdot{\bf (k+q)}}{|{\bf k}| |{\bf k+q}|}\Big] 
\frac{ - n^{F}_{{\bf k+q},\lambda^{\prime}} }{\hbar\Omega+\xi_{{\bf k},\lambda}-\xi_{{\bf k+q},\lambda^{\prime}}}, 
\nonumber\\
&=\frac{1}{2\mathcal{V}}\sum_{{\bf k}\lambda\lambda^{\prime}}
\Big[1+\lambda\lambda^{\prime} \frac{{\bf (k-q)}\cdot{\bf k}}{|{\bf k-q}| |{\bf k}|}\Big]
\frac{ n^{F}_{{\bf k},\lambda^{\prime}} }{-\hbar\Omega+\xi_{{\bf k},\lambda^{\prime}}-\xi_{{\bf k-q}\lambda}},
\end{align}
Doing similar manipulations as for $\chi^{0(+)}_{\rho\rho}({\bf q},\omega)$, the final expression of 
$\chi^{0(-)}_{\rho\rho}({\bf q},\omega)$ becomes 
 \begin{align}
\chi^{0(-)}_{\rho\rho}({\bf q},\omega)=D_{\alpha}\sum_{\lambda^{\prime}}\int_{0}^{x_{\lambda^{\prime}}^{F}} 
\frac{dx}{Q}\Big[
C_{\lambda^{\prime}}^{-}\log\Big(\frac{t_{\lambda+}^{-}-2Qx}{t_{\lambda+}^{-}+2Qx}\Big)+
G_{\lambda^{\prime}}^{-}\log\Big(\frac{t_{\lambda-}^{-}-2Qx}{t_{\lambda-}^{-}+2Qx}\Big)\Big],
\end{align}
with $\zeta_{\lambda^{\prime}}^{-}=-\hbar\Omega/\xi_{\alpha}+2\lambda^{\prime} x - Q^2$, 
$t_{\lambda^{\prime}\pm}^{-}=-(\zeta_{\lambda^{\prime}}^{-}+2) 
\pm 2\sqrt{(x+\lambda^{\prime})^2-\hbar\Omega/\xi_{\alpha}}$, $C_{\lambda^{\prime}}^{-} 
= (a_{\lambda^{\prime}}^{-} +b_{\lambda^{\prime}}^{-}t_{\lambda^{\prime}+}^{-})/
(t_{\lambda^{\prime}+}^{-}-t_{\lambda^{\prime}-}^{-})$, $G_{\lambda^{\prime}}^{-}
=-(a_{\lambda}^{-} +b_{\lambda^{\prime}}^{-}t_{\lambda^{\prime}-}^{-})/
(t_{\lambda^{\prime}+}^{-}-t_{\lambda^{\prime}-}^{-})$, $a_{\lambda^{\prime}}^{-}
=x(\zeta_{\lambda^{\prime}}^{-}+2\lambda^{\prime} x)$, and $b_{\lambda^{\prime}}^{-}
=-(\lambda^{\prime}-x)$. 
We combine $ \chi^{0(+)}_{\rho\rho}({\bf q},\omega)$ and $ \chi^{0(-)}_{\rho\rho}({\bf q},\omega)$ and 
get the following expression of the Lindhard function for $\xi_F >0 $
\begin{align}
\chi^{0}_{\rho\rho}({\bf q},\omega)=D_{\alpha}\sum_{\lambda s}\int_{0}^{x_{\lambda}^{F}} \frac{dx}{Q}\Big[
 C_{\lambda}^{s}\log\Big(\frac{t_{\lambda+}^{s}-2Qx}{t_{\lambda+}^{s}+2Qx}\Big)+
   G_{\lambda}^{s}\log\Big(\frac{t_{\lambda-}^{s}-2Qx}{t_{\lambda-}^{s}+2Qx}\Big)\Big],
\end{align}
with $s=\pm1$, $\zeta_{\lambda}^{s}=s\hbar\Omega/\xi_{\alpha}+2\lambda x - Q^2$, 
$t_{\lambda\pm}^{s}=s(\zeta_{\lambda}^{s}+2)\pm 2\sqrt{(x+\lambda)^2+s\hbar\Omega/\xi_{\alpha}}$, 
$C_{\lambda}^{s}=(a_{\lambda}^{s} +b_{\lambda}^{s}t_{\lambda+}^{s_{\omega}})/(t_{\lambda+}^{s}-t_{\lambda-}^{s})$, 
$G_{\lambda}^{s}=-(a_{\lambda}^{s} +b_{\lambda}^{s}t_{\lambda-}^{s})/(t_{\lambda+}^{s}-t_{\lambda-}^{s})$, 
$a_{\lambda}^{s}=x(\zeta_{\lambda}^{s}+2\lambda x)$, and $b_{\lambda}^{s}=s(\lambda-x)$.\\

After similar calculation the Lindhard function for $\xi_F<0$ (for $T\rightarrow 0 $) is given by
\begin{align}
\chi^{0}_{\rho\rho}({\bf q},\omega)=D_{\alpha}\sum_{s}\int_{x_{1}^{F}}^{x_{2}^{F}} \frac{dx}{Q}\Big[
C_{\lambda^{\prime}}^{s}\log\Big(\frac{t_{\lambda^{\prime}+}^{s}-2Qx}{t_{\lambda^{\prime}+}^{s}+2Qx}\Big)+
G_{\lambda^{\prime}}^{s}\log\Big(\frac{t_{\lambda^{\prime}-}^{s}-2Qx}{t_{\lambda^{\prime}-}^{s}+2Qx}\Big)\Big],
\end{align}
with $\lambda^{\prime}=-1$,, $s=\pm 1$, $x_{1}^{F}=k_{1}^{F}/k_{\alpha}$,  $x_{2}^{F}=k_{2}^{F}/k_{\alpha}$. 
While deriving the above equation, we have used the fact that $n_{{\bf k},\lambda}^{F}=0$ for all ${\bf k}$ above 
the band crossing point.

\section{Asymptotic expression of $\chi_{\rho\rho}^{0}({\bf q},\omega)$}\label{app-approx}
In this section we derive an asymptotic expression of the dynamical polarization function which will 
be helpful in finding the approximate analytical forms of plasma frequency and plasmon dispersion of 
NCMs. Let's us first consider $\xi_{F}>0$. We consider ${\bf q}=q \hat{z}$ for simplicity due to 
isotropic nature of the band structure. For small wavevector $q \ll k_{\lambda}^{F}$
\begin{align}
\xi_{{\bf k},\lambda}-\xi_{{\bf k+q},\lambda^{\prime}} & = 
\xi_{{\bf k},\lambda}-\xi_{{\bf k},\lambda^{\prime}}-{\bf q} 
\cdot {\bf \nabla}_{\bf k}\xi_{{\bf k},\lambda^{\prime}} \nonumber\\
& \simeq \alpha k (\lambda-\lambda^{\prime})- 
\hbar v_{\lambda^{\prime}}^{k} k_{\alpha}Q\cos\theta_{\bf k},
\end{align}
and for $T\rightarrow 0$
\begin{align}
n^{F}_{{\bf k},\lambda} - n^{F}_{{\bf k+q},\lambda^{\prime}} 
& = n^{F}_{{\bf k},\lambda} - n^{F}_{{\bf k},\lambda^{\prime}}- 
\frac{\partial n^{F}_{{\bf k},\lambda^{\prime}}}{\partial \xi_{{\bf k},\lambda^{\prime}}}{\bf q} 
\cdot {\bf \nabla}_{\bf k}\xi_{{\bf k},\lambda^{\prime}} \nonumber\\
& \simeq n^{F}_{{\bf k},\lambda} - n^{F}_{{\bf k},\lambda^{\prime}} 
+ \delta(\xi_{{\bf k},\lambda^{\prime}} - \xi_F)
\hbar v_{\lambda^{\prime}}^{k} k_{\alpha}Q\cos\theta_{\bf k},
\end{align}
with $v_{\lambda^{\prime}}^{k} = \hbar(k+\lambda^{\prime}k_{\alpha})/m^*$. 
So the Lindhard function will be
\begin{align}
\chi^{0}_{\rho\rho}({\bf q},\omega) & =\frac{1}{2\mathcal{V}} 
\sum_{{\bf k}\lambda\lambda^{\prime}}\Big[ 1 + \lambda\lambda^{\prime} 
\frac{{\bf k}\cdot({\bf k+q})}{|{\bf k}| |{\bf (k+q)}|}\Big] 
\frac{n^{F}_{{\bf k},\lambda} - n^{F}_{{\bf k+q},\lambda^{\prime}} }{\hbar\Omega + \xi_{{\bf k},\lambda} 
- \xi_{{\bf k+q},\lambda^{\prime}}},\nonumber\\
& \simeq \frac{1}{8\pi^2}\sum_{\lambda\lambda^{\prime}} \int_{0}^{\infty}k^2 dk\int_{0}^{\pi} 
\sin\theta_{\bf k} d\theta_{\bf k} 
\Big[1+\lambda\lambda^{\prime}(1+\frac{q}{k}\cos\theta_{\bf k})(1+\frac{q^2}{k^2} 
+ 2\frac{q}{k}\cos\theta_{\bf k})^{-1/2}\Big] \nonumber\\
&\times \frac{[ n^{F}_{{\bf k},\lambda} - n^{F}_{{\bf k},\lambda^{\prime}}
+\delta(\xi_{{\bf k},\lambda^{\prime}}-E_F)\hbar v_{\lambda^{\prime}}^{k} k_{\alpha}Q\cos\theta_{\bf k} ]}
{\hbar\Omega + \alpha k (\lambda-\lambda^{\prime}) - 
\hbar v_{\lambda^{\prime}}^{k} k_{\alpha}Q\cos\theta_{\bf k}} \nonumber\\
& = \chi^{0(3)}_{\rho\rho}({\bf q},\omega) + \chi^{0(2)}_{\rho\rho}({\bf q},\omega) 
+ \chi^{0(1)}_{\rho\rho}({\bf q},\omega).
\end{align}
Here $\chi^{0(1)}_{\rho\rho}({\bf q},\omega) + \chi^{0(2)}_{\rho\rho}({\bf q},\omega)$ is
\begin{align}
\chi^{0(2)}_{\rho\rho}({\bf q},\omega)+\chi^{0(1)}_{\rho\rho}({\bf q},\omega) 
& =\frac{1}{8\pi^2}\sum_{\lambda\lambda^{\prime}}\int_{0}^{\infty}k^2 dk\int_{0}^{\pi}\sin\theta_{\bf k} 
d\theta_{\bf k} \Big[1+\lambda\lambda^{\prime}(1+\frac{q}{k}\cos\theta_{\bf k})(1+\frac{q^2}{k^2} 
+ 2\frac{q}{k}\cos\theta_{\bf k})^{-1/2}\Big]\nonumber\\
&\times \frac{\delta(\xi_{{\bf k},\lambda^{\prime}}-\xi_F)\hbar v_{\lambda^{\prime}}^{k} 
k_{\alpha}Q\cos\theta_{\bf k} }{\hbar\Omega + \alpha k (\lambda-\lambda^{\prime})
-\hbar v_{\lambda^{\prime}}^{k} k_{\alpha}Q\cos\theta_{\bf k}}\nonumber\\
&=\frac{m k_{\alpha}}{8\pi^2\hbar^2}\sum_{\lambda\lambda^{\prime}}
\frac{(x_{\lambda^{\prime}}^{F})^2}{|x_{\lambda^{\prime}}^{F}+\lambda^{\prime} |}
\int_{-1}^{1}d\tau \Big[ 1+\lambda\lambda^{\prime}\Big(1+\frac{Q\tau}{x_{\lambda^{\prime}}^{F}}\Big)
\Big(1+\frac{Q^2}{(x_{\lambda^{\prime}}^{F})^2} + 2\frac{Q\tau}{x_{\lambda^{\prime}}^{F}}\Big)^{-1/2}\Big] \nonumber\\
&\times \gamma_{\lambda}^{\lambda^{\prime}}(x_{\lambda^{\prime}}^{F},\Omega)Q \tau 
(1-\gamma_{\lambda}^{\lambda^{\prime}}(x_{\lambda^{\prime}}^{F},\Omega) Q \tau )^{-1},
\end{align}
where $\tau = \cos \theta_{\bf k}$, $x_{\lambda^{\prime}}^{F} = k_{\lambda^{\prime}}^{F}/k_{\alpha}$, 
$Q=q/k_{\alpha}$, 
$v_{\lambda^{\prime}}^{x} = \hbar k_{\alpha}(x+\lambda^{\prime})/m^*$ and 
$\gamma_{\lambda}^{\lambda^{\prime}}(x,\Omega) = 
\hbar v_{\lambda^{\prime}}^{x} k_{\alpha}/\Delta_{\lambda}^{\lambda^{\prime}}(x,\Omega)$ with 
$\Delta_{\lambda}^{\lambda^{\prime}}(x,\Omega) = \hbar\Omega + \alpha k_{\alpha} x (\lambda- \lambda^{\prime})$. 
It is easy to see that $\gamma_{+}^{+}(x_{+}^{F},\Omega)\equiv \gamma_{+}^{+}=
\hbar v_{+}^{x_{+}^{F}} k_{\alpha}/\hbar\Omega$, $\gamma_{-}^{-}(x_{-}^{F},\Omega)\equiv\gamma_{-}^{-}
=\hbar v_{-}^{x_{-}^{F}} k_{\alpha}/\hbar\Omega$, $\gamma_{+}^{-}(x_{-}^{F},\Omega)\equiv \gamma_{+}^{-}
=\hbar v_{-}^{x_{-}^{F}} k_{\alpha}/(\hbar\Omega+2\alpha k_{\alpha} x_{-}^{F} )$ and 
$\gamma_{-}^{+}(x_{+}^{F},\Omega)\equiv \gamma_{-}^{+}=
\hbar v_{+}^{x_{+}^{F}} k_{\alpha}/(\hbar\Omega-2\alpha k_{\alpha} x_{+}^{F} )$ with 
$\gamma_{+}^{+} = \gamma_{-}^{-}$ as the velocities at the two bands are the same for 
a given Fermi energy. 
So the final expression of intraband contribution $\chi^{0(1)}_{\rho\rho}({\bf q},\omega)$ becomes
\begin{align}\label{chi0pintra}
\chi^{0(1)}_{\rho\rho}({\bf q},\omega)&=D_{\alpha}
\sum_{\lambda=\lambda^{\prime}}\frac{(x_{\lambda^{\prime}}^{F})^2}{|x_{\lambda^{\prime}}^{F}+\lambda^{\prime} |}
\Big[ (1+\lambda\lambda^{\prime})\frac{1}{3}(Q\gamma_{\lambda}^{\lambda^{\prime}}(x_{\lambda^{\prime}}^{F},\Omega))^2 + 
\lambda\lambda^{\prime}\Big(\frac{2\gamma_{\lambda}^{\lambda^{\prime}}(x_{\lambda^{\prime}}^{F},\Omega)}
{15 (x_{\lambda^{\prime}}^{F})^3}-
\frac{(\gamma_{\lambda}^{\lambda^{\prime}}(x_{\lambda^{\prime}}^{F},\Omega))^2}
{15 (x_{\lambda^{\prime}}^{F})^2}\Big)Q^4\nonumber\\
&+(1+\lambda\lambda^{\prime})\frac{1}{5}(Q\gamma_{\lambda}^{\lambda^{\prime}}(x_{\lambda^{\prime}}^{F},\Omega))^4 +
\lambda\lambda^{\prime}\Big(-\frac{2(\gamma_{\lambda}^{\lambda^{\prime}}(x_{\lambda^{\prime}}^{F},\Omega))^2}
{35 (x_{\lambda^{\prime}}^{F})^4} + \frac{2(\gamma_{\lambda}^{\lambda^{\prime}}(x_{\lambda^{\prime}}^{F},\Omega))^3}
{35 (x_{\lambda^{\prime}}^{F})^3}-
\frac{(\gamma_{\lambda}^{\lambda^{\prime}}(x_{\lambda^{\prime}}^{F},\Omega))^4}
{35 (x_{\lambda^{\prime}}^{F})^2}\Big)Q^6\nonumber\\
&+(1+\lambda\lambda^{\prime})\frac{1}{7}(Q\gamma_{\lambda}^{\lambda^{\prime}}(x_{\lambda^{\prime}}^{F},\Omega))^6 
+ \mathcal{O}(Q^8)+...\Big] ,
\end{align}
and the final expression of one part of interband contribution 
$\chi^{0(2)}_{\rho\rho}({\bf q},\omega)$ becomes
\begin{align}\label{chi0pinter1}
\chi^{0(2)}_{\rho\rho}({\bf q},\omega)&=D_{\alpha}\frac{(x_{-}^{F})^2}{|x_{-}^{F}-1 |}\Big[  
\Big(-\frac{2\gamma_{+}^{-}(x_{-}^{F},\Omega)}{15 (x_{-}^{F})^3} +\frac{(\gamma_{+}^{-}(x_{-}^{F},\Omega))^2}
{15 (x_{-}^{F})^2}\Big)Q^4\nonumber\\
&+ \Big(\frac{2(\gamma_{+}^{-}(x_{-}^{F},\Omega))^2}{35 (x_{-}^{F})^4} - 
\frac{2(\gamma_{+}^{-}(x_{-}^{F},\Omega))^3}{35(x_{-}^{F})^3}+
\frac{(\gamma_{+}^{-}(x_{-}^{F},\Omega))^4}{35 (x_{-}^{F})^2}\Big)Q^6
+ \mathcal{O}(Q^8)+...\Big] \nonumber\\
&+D_{\alpha}\frac{(x_{+}^{F})^2}{|x_{+}^{F}+1 |}\Big[  \Big(-\frac{2\gamma_{-}^{+}(x_{+}^{F},\Omega)}
{15 (x_{+}^{F})^3} +\frac{(\gamma_{-}^{+}(x_{+}^{F},\Omega))^2}{15 (x_{+}^{F})^2}\Big)Q^4\nonumber\\
&+ \Big(\frac{2(\gamma_{-}^{+}(x_{+}^{F},\Omega))^2}{35 (x_{+}^{F})^4} - \frac{2(\gamma_{-}^{+}(x_{+}^{F},\Omega))^3}
{35(x_{+}^{F})^3}+
\frac{(\gamma_{-}^{+}(x_{+}^{F},\Omega))^4}{35 (x_{+}^{F})^2}\Big)Q^6+ \mathcal{O}(Q^8)+...\Big] .
\end{align}
The remaining part of the interband contribution $\chi^{0(3)}_{\rho\rho}({\bf q},\omega)$ is finally given by
\begin{align}\label{chi0pinter2}
\chi^{0(3)}_{\rho\rho}({\bf q},\omega)&=D_{\alpha}\Big[\frac{\xi_{\alpha}Q^2}{3\alpha k_{\alpha}}
\Big[\log\Big(\frac{(\hbar\Omega)^2 - (2\alpha k_{+}^F)^2}{(\hbar\Omega)^2 - (2\alpha k_{-}^F)^2}\Big) \Big]
+ \frac{8\xi_{\alpha}^2Q^4}{15}\Big[\frac{1}{(\hbar\Omega)^2}\log\Big(\frac{4-(\hbar\Omega/(\alpha k_{+}^F))^2}
{4-(\hbar\Omega/(\alpha k_{-}^F))^2}\Big) \nonumber\\
&-\frac{4\alpha k_{\alpha}(x_{-}^F - x_{+}^F)(x_{-}^F + x_{+}^F)(\hbar\Omega+2\alpha k_{\alpha})}
{((2\alpha k_{-}^F)^2-(\hbar\Omega)^2)((2\alpha k_{+}^F)^2-(\hbar\Omega)^2)}\Big]
+ \frac{8\xi_{\alpha}^3Q^4}{15}\Big[\frac{1}{(2\alpha k_{\alpha})^3}
\log\Big(\frac{(\hbar\Omega)^2 - (2\alpha k_{+}^F)^2}{(\hbar\Omega)^2 - (2\alpha k_{-}^F)^2}\Big) \nonumber\\
&-\frac{1}{[(2\alpha k_{\alpha})^3((\hbar\Omega)^2-(2\alpha k_{-}^F)^2)^2((\hbar\Omega)^2-(2\alpha k_{+}^F)^2)^2]}
[4\alpha^2 k_{\alpha}^2(x_{-}^F - x_{+}^F)(x_{-}^F + x_{+}^F)(\hbar\Omega+2\alpha k_{\alpha})\nonumber\\
&\times[ 32\alpha^5k_{\alpha}^5(x_{-}^Fx_{+}^F)^2 + 
80\alpha^4 k_{\alpha}^4\hbar\Omega(x_{-}^Fx_{+}^F)^2 + 
4\alpha^2 k_{\alpha}^2(\hbar\omega)^2((x_{+}^F)^2+(x_{-}^F)^2)(2\alpha k_{\alpha}-3\hbar\Omega)
-6\alpha k_{\alpha}(\hbar\Omega)^4 + (\hbar\Omega)^5]]\Big]\nonumber\\
&+\mathcal{O}(Q^6) +...\Big].
\end{align}
Equations.~\ref{chi0pintra},~\ref{chi0pinter1} and \ref{chi0pinter2} combinedly describe the asymptotic 
expression of the Lindhard function for $\xi_F>0$.\\

Now we consider the Lindhard function for $\xi_F<0$,
\begin{align}
\chi^{0}_{\rho\rho}({\bf q},\omega)&=\frac{1}{2\mathcal{V}}\sum_{{\bf k}}
\Big[1+\frac{{\bf k}\cdot({\bf k+q})}{|{\bf k}| |{\bf k+q}|}\Big]
\frac{n^{F}_{{\bf k},-} - n^{F}_{{\bf k+q},-} }{\hbar\Omega+\xi_{{\bf k},-}-\xi_{{\bf k+q},-}}\nonumber\\
&+\frac{1}{2\mathcal{V}}\sum_{{\bf k}}\Big[1-\frac{{\bf k}\cdot({\bf k+q})}{|{\bf k}| |{\bf k+q}|}\Big]
\Big[\frac{n^{F}_{{\bf k},-}}{\hbar\Omega+\xi_{{\bf k},-}-\xi_{{\bf k+q},+}} - \frac{n^{F}_{{\bf k+q},-}}
{\hbar\Omega+\xi_{{\bf k},+}-\xi_{{\bf k+q},-}}\Big],\nonumber\\
&=\frac{1}{2\mathcal{V}}\sum_{{\bf k}}\Big[1+\frac{{\bf k}\cdot({\bf k+q})}{|{\bf k}| |{\bf k+q}|}\Big]
\frac{\delta(\xi_{{\bf k},-}-\xi_F)\hbar v_{-}^{k} k_{\alpha}Q\cos\theta_{\bf k} }
{\hbar\Omega-\hbar v_{-}^{k} k_{\alpha}Q\cos\theta_{\bf k}}\nonumber\\
&+\frac{1}{2\mathcal{V}}\sum_{{\bf k}}\Big[1-\frac{{\bf k}\cdot({\bf k+q})}{|{\bf k}| |{\bf k+q}|}\Big]
\frac{\delta(\xi_{{\bf k},-}-\xi_F)\hbar v_{-}^{k} k_{\alpha}Q\cos\theta_{\bf k} }
{\hbar\Omega+2\alpha k-\hbar v_{-}^{k} k_{\alpha}Q\cos\theta_{\bf k}}\nonumber\\
&+\frac{1}{2\mathcal{V}}\sum_{{\bf k}}\Big[1-\frac{{\bf k}\cdot({\bf k+q})}{|{\bf k}| |{\bf k+q}|}\Big]
\Big[\frac{n^{F}_{{\bf k},-}}{\hbar\Omega-2\alpha k-\hbar v_{+}^{k} k_{\alpha}Q\cos\theta_{\bf k}} - 
\frac{n^{F}_{{\bf k},-}}{\hbar\Omega+2\alpha k-\hbar v_{-}^{k} k_{\alpha}Q\cos\theta_{\bf k}}\Big],\nonumber\\
&=\chi^{0(1)}_{\rho\rho}({\bf q},\omega)+\chi^{0(2)}_{\rho\rho}({\bf q},\omega) 
+ \chi^{0(3)}_{\rho\rho}({\bf q},\omega).
\end{align}
Following the similar steps as in $\xi_F>0$, the final expression for intraband $\lambda=\pm 1$ or intrabranch 
$\eta=1,2$ contribution to the Lindhard function is given by
\begin{align}\label{chi0mintra}
\chi^{0(1)}_{\rho\rho}({\bf q},\omega)
&=D_{\alpha}\sum_{\eta}\frac{(x_{\eta}^{F})^2}{|x_{\eta}^{F}- 1 |}
\Big[ \frac{2}{3}(Q\gamma_{-}^{-}(x_{\eta}^{F},\Omega))^2 + 
\Big(\frac{2\gamma_{-}^{-}(x_{\eta}^{F},\Omega)}{15 (x_{\eta}^{F})^3} 
-\frac{(\gamma_{-}^{-}(x_{\eta}^{F},\Omega))^2}{15 (x_{\eta}^{F})^2}\Big)Q^4\nonumber\\
&+\frac{2}{5}(Q\gamma_{-}^{-}(x_{\eta}^{F},\Omega))^4 + 
\Big(-\frac{2(\gamma_{-}^{-}(x_{\eta}^{F},\Omega))^2}{35 (x_{\eta}^{F})^4} 
+ \frac{2(\gamma_{-}^{-}(x_{\eta}^{F},\Omega))^3}{35 (x_{\eta}^{F})^3}
-\frac{(\gamma_{-}^{-}(x_{\eta}^{F},\Omega))^4}{35 (x_{\eta}^{F})^2}\Big)\mathcal{Q}^6\nonumber\\
&+ \frac{2}{7}(Q\gamma_{-}^{-}(x_{\eta}^{F},\Omega))^6 + \mathcal{O}(Q^8)+...\Big].
\end{align}
with $x_{\eta}^F = k_{\eta}^{F}/k_{\alpha}$,  
and $\gamma_{-}^{-}(x_{\eta}^{F},\Omega) = \xi_{\alpha}(x_{\eta}^F -1)/(\hbar\Omega)$. 
The final expression of one part of interband and intrabranch contribution becomes
\begin{align}\label{chi0minter1}
\chi^{0(2)}_{\rho\rho}({\bf q},\omega)
&=D_{\alpha}\sum_{\eta}\frac{(x_{\eta}^{F})^2}{|x_{\eta}^{F}- 1 |}
\Big[ \Big(-\frac{2\gamma_{+}^{-}(x_{\eta}^{F},\Omega)}{15 (x_{\eta}^{F})^3} + 
\frac{(\gamma_{+}^{-}(x_{\eta}^{F},\Omega))^2}{15 (x_{\eta}^{F})^2}\Big)\mathcal{Q}^4\nonumber\\
&+ \Big(\frac{2(\gamma_{+}^{-}(x_{\eta}^{F},\Omega))^2}{35 (x_{\eta}^{F})^4} - 
\frac{2(\gamma_{+}^{-}(x_{\eta}^{F},\Omega))^3}{35 (x_{\eta}^{F})^3}+
\frac{(\gamma_{+}^{-}(x_{\eta}^{F},\Omega))^4}{35 (x_{\eta}^{F})^2}\Big)Q^6+ \mathcal{O}(Q^8)+...\Big].
\end{align}
with $x_{\eta}^F = k_{\eta}^{F}/k_{\alpha}$
and $\gamma_{+}^{-}(x_{\eta}^{F},\Omega) = 
\xi_{\alpha}(x_{\eta}^F -1)/(\hbar\Omega+2\alpha k_{\alpha}x_{\eta}^F)$. 
The remaining part of the intraband and intrabranch contribution to the Lindhard function has similar 
expression as that of $\chi_{\rho\rho}^{0(3)}({\bf q},\omega)$ for $\xi_F>0$ except $k_{+}^F$ is replaced 
$k_{1}^F$ and $k_{-}^F$ by $k_{2}^F$. So the final expression of $\chi_{\rho\rho}^{0(3)}({\bf q},\omega)$ 
for $\xi_F<0$ becomes
\begin{align}\label{chi0minter2}
\chi^{0(3)}_{\rho\rho}({\bf q},\omega)&=D_{\alpha}\Big[\frac{\xi_{\alpha}Q^2}{3\alpha k_{\alpha}}
\Big[\log\Big(\frac{(\hbar\Omega)^2 - (2\alpha k_{1}^F)^2}{(\hbar\Omega)^2 - (2\alpha k_{2}^F)^2}\Big) \Big] + 
\frac{8\xi_{\alpha}^2Q^4}{15}\Big[\frac{1}{(\hbar\Omega)^2}\log\Big(\frac{4-(\hbar\Omega/(\alpha k_{1}^F))^2}
{4-(\hbar\Omega/(\alpha k_{2}^F))^2}\Big) \nonumber\\
&-\frac{4\alpha k_{\alpha}(x_{2}^F - x_{1}^F)(x_{2}^F + x_{1}^F)
(\hbar\Omega+2\alpha k_{\alpha})}{((2\alpha k_{2}^F)^2-(\hbar\Omega)^2)((2\alpha k_{1}^F)^2-
(\hbar\Omega)^2)}\Big]+ \frac{8\xi_{\alpha}^3Q^4}{15}\Big[\frac{1}{(2\alpha k_{\alpha})^3}
\log\Big(\frac{(\hbar\Omega)^2 - (2\alpha k_{1}^F)^2}{(\hbar\Omega)^2 - (2\alpha k_{2}^F)^2}\Big) \nonumber\\
&-\frac{1}{[(2\alpha k_{\alpha})^3((\hbar\Omega)^2-(2\alpha k_{2}^F)^2)^2((\hbar\Omega)^2-(2\alpha k_{1}^F)^2)^2] }
[4\alpha^2 k_{\alpha}^2(x_{2}^F - x_{1}^F)(x_{2}^F + x_{1}^F)(\hbar\Omega+2\alpha k_{\alpha})\nonumber\\
&\times[ 32\alpha^5k_{\alpha}^5(x_{2}^Fx_{1}^F)^2 + 80\alpha^4 k_{\alpha}^4\hbar\Omega(x_{2}^Fx_{1}^F)^2 + 
4\alpha^2 k_{\alpha}^2(\hbar\omega)^2((x_{2}^F)^2+(x_{1}^F)^2)(2\alpha k_{\alpha}-3\hbar\Omega)-
6\alpha k_{\alpha}(\hbar\Omega)^4 + (\hbar\Omega)^5]]\Big]\nonumber\\
& + \mathcal{O}(Q^6) +...\Big].
\end{align}
The full asymptotic expression of the Lindhard function for $\xi_F<0$ is the sum 
of Eqs.~\ref{chi0mintra},~\ref{chi0minter1} and \ref{chi0minter2}.

\section{Density-density response in presence of electron-electron interaction}\label{app-plasmon}
The  Coulomb interaction among the band electrons in second quantized form can be written as follows\cite{giuliani,bruusflensberg}
\begin{align}
\hat{V}=\frac{1}{2}\sum_{\sigma, \sigma^{\prime}}\int d{\bf r}
\int d{\bf r}^{\prime}\tilde{\Psi}^{\dagger}_{\sigma}({\bf r})\tilde{\Psi}^{\dagger}_{\sigma^{\prime}}({\bf r}^{\prime})
\frac{e_{0}^{2}}{|{\bf r}^{\prime}-{\bf r}|}\tilde{\Psi}_{\sigma^{\prime}}({\bf r}^{\prime})\tilde{\Psi}_{\sigma}({\bf r}),
\end{align}
where $e_{0}^{2}=e^2/(4\pi\epsilon)$, with $\epsilon$ being the background dielectric constant. 
After following the well known procedure within jellium model, the electron-electron interaction in second 
quantized form takes the following form in the helicity basis as
\begin{align}
\hat{V}=\frac{1}{2\mathcal{V}}\sum_{\substack{{\bf k}_{1},{\bf k}_2 , {\bf q}^{\prime}\neq 0\\ 
\lambda_1,\lambda_2,\lambda_3,\lambda_4}} V( q^{\prime})C^{\dagger}_{{\bf k}_1 + {\bf q}^{\prime},\lambda_1}
\phi^{\dagger}_{{\bf k}_1 + {\bf q}^{\prime},\lambda_1}C^{\dagger}_{{\bf k}_2-{\bf q}^{\prime},\lambda_2}
\phi^{\dagger}_{{\bf k}_2-{\bf q}^{\prime},\lambda_2}
\phi_{{\bf k}_2,\lambda_3}C_{{\bf k}_2,\lambda_3}\phi_{{\bf k}_1,\lambda_4}
C_{{\bf k}_1,\lambda_4},
\end{align}
with $ V( q^{\prime}) = 4\pi e_{0}^{2}/(q^{\prime})^2$. 
In presence of the elctron-electron interaction the induced particle density 
due to the external perturbation defined above becomes
\begin{align}
\rho_{\rm ind}^{i}({\bf r},t)=\int_{-\infty}^{t} d t^{\prime} \int d{\bf r}^{\prime} 
\chi^{i}_{\rho\rho}({\bf r},{\bf r}^{\prime},t,t^{\prime})V_{\rm ext}({\bf r}^{\prime},t^{\prime}),
\end{align} 
where $\chi^{i}_{\rho\rho}({\bf r},{\bf r}^{\prime},t,t^{\prime})$ is the retarded density-density response 
function for the system described by the total Hamiltonian $\hat{H}=\hat{H}_0 + \hat{V}$ and has following form
\begin{align}
\chi^{i}_{\rho\rho}({\bf r},{\bf r}^{\prime},t,t^{\prime})=-\frac{i}{\hbar}\theta(t-t^{\prime})
\langle[\hat{\rho}({\bf r},t),\hat{\rho}({\bf r}^{\prime},t^{\prime})]\rangle_{\rm eq}.
\end{align} 
Here the subscript $'{\rm eq}'$ denotes that the average is taken over the ground state of the full 
Hamiltonian $\hat{H}=\hat{H}_0 + \hat{V}$ in equilibrium. Using the properties of the 
translationally invariant system even in the presence of the Coulomb interaction 
the density response function takes the following form
\begin{align}
\chi^{i}_{\rho\rho}({\bf q},t,t^{\prime})=\sum_{\lambda\lambda^{\prime}}\chi_{\lambda\lambda^{\prime}}^{i}
({\bf q},t,t^{\prime})=-\frac{i}{\hbar\mathcal{V}}\Theta(t-t^{\prime})
\langle[\hat{\rho}({\bf q},t),\hat{\rho}(-{\bf q},t^{\prime})]\rangle_{\rm eq}.
\end{align} 
We use the standard equation of motion technique within the random phase approximation
to obtain the final expression of the density-density response function of the interacting 
system which is given by\cite{giuliani,bruusflensberg}
\begin{align}
\chi^{i}_{\rho\rho}({\bf q},\omega)=\sum_{\lambda\lambda^{\prime}}\chi_{\lambda\lambda^{\prime}}^{i}
({\bf q},\omega)=\frac{\chi_{\rho\rho}^{0}({\bf q},\omega)}{1-V({\bf q})\chi_{\rho\rho}^{0}({\bf q},\omega)},
\end{align} 
where $\chi_{\rho\rho}^{0}({\bf q},\omega)$ is described by Eq.~\ref{chi0A}. The plasmons are described by
the poles of the above response function $\textit{i.e.}$ zeros of the dielectric function 
$\epsilon({\bf q},\omega) = 1-V({\bf q})\chi_{\rho\rho}^{0}({\bf q},\omega)$.
 
\section{Optical Conductivity} \label{app-op-con}
Let's first consider the NCMs without electron-electron interaction in presence of an external 
perturbation $\hat{V}_{\rm ext}(t)= \int d{\bf r} V_{\rm ext}({\bf r},t)\hat{\rho}({\bf r})$. 
The induced density due to this perturbation in Fourier space is given by\cite{bruusflensberg}
\begin{align}
\rho_{\rm ind}({\bf q},\omega)=\chi^{0}_{\rho\rho}({\bf q},\omega)V_{\rm ext}({\bf q},\omega),
\end{align} 
with $\chi^{0}_{\rho\rho}({\bf q},\omega)$ being the retarded density-density response for noninteracting 
NCMs. The continuity equation $\partial_{t} \rho_{\rm ind}({\bf r},t) + {\bs \nabla}\cdot{\bf J}({\bf r},t) = 0$ 
in Fourier space becomes $-i\omega \rho_{\rm ind}({\bf q},\omega) + i{\bf q}\cdot{\bf J} ({\bf q},\omega)=0$, 
with the electrical current ${\bf J} ({\bf q},\omega)=\sigma({\bf q},\omega){\bf E}_{\rm ext}({\bf q},\omega)$ in 
presence of an external electric field ${\bf E}_{\rm ext}({\bf q},\omega) = -i{\bf q}V_{\rm ext}({\bf q},\omega)$. 
With the help of the above relations the relation between the longitudinal conductivity $\sigma({\bf q},\omega)$ 
and the dynamical polarization function $\chi^{0}_{\rho\rho}({\bf q},\omega)$ is given by
\begin{align}
\sigma({\bf q},\omega)=\frac{i\omega e^2}{q^2}\chi^{0}_{\rho\rho}({\bf q},\omega).
\end{align}
In presence of the electron-electron interaction, the above equations modifies as follows
\begin{align}
\rho_{\rm ind}^{i}({\bf q},\omega)=\chi^{i}_{\rho\rho}({\bf q},\omega)V_{\rm ext}({\bf q},\omega),
\end{align} 
where $\chi^{i}_{\rho\rho}({\bf q},\omega)$ is the density response function for 
$\hat{H} = \hat{H}_0 +\hat{V}$ with induced particle density $\rho_{\rm ind}^{i}({\bf q},\omega)$ within RPA. 
The continuity equation is also modified in a same way as 
$-i\omega \rho_{\rm ind}^{i}({\bf q},\omega) + i{\bf q}\cdot{\bf J}^{i} ({\bf q},\omega)=0$, 
giving rise to the following relation between $\sigma^{i}({\bf q},\omega)$ and $\chi^{i}_{\rho\rho}({\bf q},\omega)$ 
\begin{align}
\sigma^{i}({\bf q},\omega)=\frac{i\omega e^2}{q^2}\chi^{i}_{\rho\rho}({\bf q},\omega).
\end{align}
\end{widetext}


\begin{thebibliography}{55}


\bibitem{Rash1}
        E. I. Rashba,
        Sov. Phys. Solid State {\bf 2}, 1109 (1960).

        \bibitem{Rash2}
        Y. A. Bychkov and E. I. Rashba,
        J. Phys. C {\bf 17}, 6039 (1984).

        \bibitem{Dress}
        G. Dresselhaus,
        Phys. Rev. {\bf 100}, 580 (1955).


	
	\bibitem{spintro1}
	S. A. Wolf, D. D. Awschalom, R. A. Buhrman, J. M. Daughton, S. von. Molnar, 
	M. L. Roukes, A. Y. Chtchelkanova, and D. M Treger,
	Science {\bf 294}, 1488 (2001).
	
	
	\bibitem{spintro2}
	R. Winkler,
	{\it Spin-Orbit Coupling Effets in Two-dimensional Electron and Hole
	Systems (Springer Berlin Heidelberg, 2003)}.
	
	
	\bibitem{spintro3}
	I. Zutic, J. Fabian, and S. Das Sarma,
	Rev. Mod. Phys. {\bf 76}, 323 (2004)


        \bibitem{dissp1}
        S. Murakami, N. Nagaosa, and S. C. Zhang,
        Science {\bf 301}, 1348 (2003)

        \bibitem{dissp2}
        S. Murakami, N. Nagaosa, and S. C. Zhang,
        Phys. Rev. B {\bf 69}, 235206 (2004)

        \bibitem{SHE1}
        J. E. Hirsch,
        Phys. Rev. Lett. {\bf 83}, 1834 (1999).

        \bibitem{SHE2}
        S. Zhang,
        Phys. Rev. Lett. {\bf 85}, 393 (2000).

        \bibitem{SHE3}
        J. Sinova, D. Culcer, Q. Niu, N. A. Sinitsyn, T. Jungwirth, and A. H. MacDonalds,
        Phys. Rev. Lett. {\bf 92}, 126603 (2004).


        \bibitem{SHE4}
        Y. K. Kato, R. C. Myers, A. C. Gossard, and D. D. Awschalom,
        Science {\bf 306}, 1910 (2004)

        \bibitem{SHE5}
        B. A. Bernevig and S. C. Zhang,
        Phys. Rev. Lett. {\bf 96}, 106802 (2006).

        \bibitem{SHE6}
        J. Sinova, S. O. Valenzuela, J. Wunderlich, C. H. Back, and T. Jungwirth,
        Rev. Mod. Phys. {\bf 87}, 1213 (2015).

        \bibitem{spin-torque}
        K. Tsutsui and S. Murakami,
        Phys. Rev. B {\bf 86}, 115201 (2012).

        \bibitem{SGE1}
        S. D. Ganichev, E. L. Ivchenko, V. V. Belkov, S. A. Tarasenko, M. Sollinger,
        D. Weiss, W. Wegscheider, and W. Prettl,
        Nature {\bf 417}, 153 (2002).

		\bibitem{giuliani}
		Gabriele Giuliani and Giovanni Vignale, {\it Quantum Theory of the Electron Liquid} (Cambridge University Press, Cambridge, 2005).
		
		\bibitem{bruusflensberg}
		Henrik Bruus and Karsten Flensberg, {\it Many-Body Quantum Theory in Condensed Matter Physics: An Introduction} (Oxford University Press, Oxford, 2004)
	
	
%
	
	\bibitem{2dplasmon1}
	G-H Chen and M. E. Raikh, Phys. Rev. B {\bf 59}, 5090 (1999).
	
	\bibitem{2dplasmon3}
	M. Pletyukhov and V. Gritsev,	Phys. Rev. B {\bf 74}, 045307 (2006).
	
	\bibitem{metaldichalconide}
	D. K. Mukherjee, A. Kundu, and H. A. Fertig, Phys. Rev. B {\bf 98}, 184413 (2018).
	
	\bibitem{2dplasmon4}
	S. M. Badalyan, A. Matos-Abiague, G. Vignale, and J. Fabian, Phys. Rev. B {\bf 79}, 205305 (2009).
	
       \bibitem{2DHGanisotropic}
		A. Scholz, T. Dollinger, P. Wenk, K. Richter, and J. Schliemann, Phys. Rev. B {\bf 87}, 085321 (2013).
	
	\bibitem{2dplasmon2}
	X. F. Wang, Phys. Rev. B {\bf 72}, 085317 (2005).
	
	
	
	
	
	
	
	
	
	
	
	
	





	
	\bibitem{BiSe1}
	M. Bianchi, D. Guan, S. Bao, J. Mi, B. B. Iversen, P. D.C. King, and P. Hofmann,
	Nature Communications {\bf 1}, 128 (2010).
	
	
	\bibitem{BiSe2}
	P. D. C. King, R. C. Hatch, M. Bianchi, R. Ovsyannikov, C. Lupulescu, G. Landolt, 
	B. Slomski, J. H. Dil, D. Guan, J. L. Mi, E. D. L. Rienks, J. Fink, A. Lindblad, 
	S. Svensson, S. Bao, G. Balakrishnan, B. B. Iversen, J. Osterwalder, W. Eberhardt, 
	F. Baumberger, and Ph. Hofmann,
	Phys. Rev. Lett. {\bf 107}, 096802 (2011).
	
	\bibitem{Bi_Alloy}
	C. R. Ast, J. Henk, A. Ernst, L. Moreschini, M. C. Falub,
	D. Pacile, P. Bruno, K. Kern, and M. Grioni,
	Phys. Rev. Lett. {\bf 98}, 186807 (2007).
	
	
	\bibitem{BiTeI1}
	K. Ishizaka, M. S. Bahramy, H. Murakawa, M. Sakano, T. Shimojima, T. Sonobe, K. Koizumi,
	S. Shin, H. Miyahara, A. Kimura, K. Miyamoto, T. Okuda, H. Namatame, M. Taniguchi,
	R. Arita, N. Nagaosa, K. Kobayashi, Y. Murakami, R. Kumai, Y. Kaneko, Y. Onose, and Y. Tokura,
	Nat. Mater. {\bf 10}, 521 (2011). 
	
	\bibitem{BiTeI2}
	M. S. Bahramy, R. Arita, and N. Nagaosa,
	Phys. Rev. B {\bf 84}, 041202(R) (2011).
	
	\bibitem{BiTeI3}
	S. V. Eremeev, I. A. Nechaev, Yu. M. Koroteev, P. M. Echenique, and E. V. Chulkov,
	Phys. Rev. Lett. {\bf 108}, 246802 (2012).
	
	\bibitem{BiTeI4}
	G. Landolt, S. V. Eremeev, Y. M. Koroteev, B. Slomski, S. Muff, T. Neupert,
	M. Kobayashi, V. N. Strocov, T. Schmitt, Z. S. Aliev, M. B. Babanly,
	I. R. Amiraslanov, E. V. Chulkov, J. Osterwalder, and J. H. Dil,
	Phys. Rev. Lett. {\bf 109}, 116403 (2012).
	
	
	\bibitem{BiTeI5}
	M. Sakano, M. S. Bahramy, A. Katayama, T. Shimojima, H. Murakawa, Y. Kaneko, W. Malaeb, 
	S. Shin, K. Ono, H. Kumigashira, R. Arita, N. Nagaosa, H. Y. Hwang, Y. Tokura, and K. Ishizaka,
	Phys. Rev. Lett. {\bf 110}, 107204 (2013).
	

        \bibitem{B201}
        J. Kang and J. Zang,
        Phys. Rev. B {\bf 91}, 134401 (2015).

        \bibitem{NC_Li1}
        V. P. Mineev and Y. Yoshioka,
        Phys. Rev. B {\bf 81}, 094525 (2010).


        \bibitem{susRashba}
        I. I. Boiko and E. I. Rashba, Fiz. Tverd. Tela (Leningrad)
        {\bf 2}, 1874 (1960) [Sov. Phys. Solid State {\bf 2}, 1692 (1960)].

        \bibitem{susprl}
        G. A. H. Schober, H. Murakawa, M. S. Bahramy, R. Arita, Y. Kaneko, Y. Tokura,  and N. Nagaosa,
        Phys. Rev. Lett. {\bf 108}, 247208 (2012).


	\bibitem{Trans1}
	H. Murakawa, M. S. Bahramy, M. Tokunaga, Y. Kohama, C.
	Bell, Y. Kaneko, N. Nagaosa, H. Y. Hwang, and Y. Tokura,
	Science {\bf 342}, 1490 (2013).
	
	\bibitem{Trans2}
	C. Bell, M. S. Bahramy, H. Murakawa, J. G. Checkelsky,
	R. Arita, Y. Kaneko, Y. Onose, M. Tokunaga, Y. Kohama,
	N. Nagaosa, Y. Tokura, and H. Y. Hwang, 
	Phys. Rev. B {\bf 87}, 081109(R) (2013).
	
	\bibitem{Trans3}
	C. Martin, E. D. Mun, H. Berger, V. S. Zapf, and D. B. Tanner,
	Phys. Rev. B {\bf 87}, 041104(R) (2013).
	
	\bibitem{Trans4}
	T. Ideue, J. G. Checkelsky, M. S. Bahramy, H. Murakawa,
	Y. Kaneko, N. Nagaosa, and Y. Tokura, 
	Phys. Rev. B {\bf 90}, 161107(R) (2014).
	
	\bibitem{Trans5}
	L. Ye, J. G. Checkelsky, F. Kagawa, and Y. Tokura, 
	Phys. Rev. B {\bf 91}, 201104(R) (2015).
	
	\bibitem{Trans6}
	C. R. Wang, J. C. Tung, R. Sankar, C. T. Hsieh, Y. Y. Chien,
	G. Y. Guo, F. C. Chou, and W. L. Lee, 
	Phys. Rev. B {\bf 88}, 081104(R) (2013).
	
	\bibitem{Trans7}
	V. Brosco and C. Grimaldi,
	Phys. Rev. B {\bf 95}, 195164 (2017).
	
	\bibitem{Mag_pht}
	N. Ogawa, M. S. Bahramy, H. Murakawa, Y. Kaneko, and Y. Tokura, 
	Phys. Rev. B {\bf 88}, 035130 (2013).
	
	\bibitem{Opt3}
	S. Maiti, V. Zyuzin, and D. L. Maslov,
	Phys. Rev. B {\bf 91}, 035106 (2015).
	
	\bibitem{op-cond-ti}
	P. Di Pietro, F. M. Vitucci, D. Nicoletti, L. Baldassarre,
	P. Calvani, R. Cava, Y. S. Hor, U. Schade, and S. Lupi,
	Phys. Rev. B {\bf 86}, 045439 (2012). 
	
	
	\bibitem{Therm1}
	L. Wu, J. Yang, S. Wang, P. Wei, J. Yang, W. Zhang, and L. Chen,
	Phys. Rev. B {\bf 90}, 195210 (2014).
	
	\bibitem{Therm2}
	T. Ideue, L. Ye, J. G. Checkelsky,  H. Murakawa, Y. Kaneko, and Y. Tokura,
	Phys. Rev. B {\bf 92}, 115144 (2015).
	

       \bibitem{RKKY}
        S. X. Wang, H. R. Chang, and J. Zhou,
        Phys. Rev. B {\bf 96}, 115204 (2017).







	
	
	
	

	
	
	
	
	
	\bibitem{sup1}
	K. Togano, P. Badica, Y. Nakamori, S. Orimo, H. Takeya, and K. Hirata,
	Phys. Rev. Lett. {\bf 93}, 247004 (2004).
	
	\bibitem{rev-sup}
	M. Smidman, M. B. Salamon, H. Q. Yuan, and D. F. Agterber,
	Rep. Prog. Phys. {\bf 80}, 036501 (2017).
	

	\bibitem{NC_Li2}
	K. V. Samokhin, 
	Phys. Rev. B {\bf 78}, 144511 (2008).
	
	\bibitem{NC_Li3}
	V. P. Mineev,
	Phys. Rev. B {\bf 88}, 134514 (2013).
	
	
	\bibitem{spin_scp}
	K. V. Samokhin, 
	Phys. Rev. B {\bf 76}, 094516 (2007).
	
	
	\bibitem{thermNCMs}
	S. Verma, T. Biswas, and T. K. Ghosh, Phys. Rev. B {\bf 100}, 045201 (2019).


        
	
		
        \bibitem{fullbandstruc}
        K.-W. Lee and W. E. Pickett,
        Phys. Rev. B \textbf{72}, 174505 (2005).


	\bibitem{singularbilayer}
	S. Dey and R. Sensarma, Phys. Rev. B {\bf 94}, 235107 (2016).
	
	\bibitem{epsilonBiTeI}
X. Xi, C. Ma, Z. Liu, Z. Chen, W. Ku, H. Berger, C. Martin, D. B. Tanner, and G. L. Carr, Phys. Rev. Lett. {\bf 111}, 155701 (2013).

	
	
	
	
	
	
	
	
	
	
	
	
\end{thebibliography}
\end{document}